# Emergence of millimeter-wave resonances in self-assembled ferroelectric metamaterials

Florian Bergmann[1,2*], Peter Meisenheimer[3*], Aiden Ross[4*], Marvin Schewe[5], Fernando Gómez-Ortiz[6], Kaiwen Yang[7], Xinyan Li[8,9], Thomas J. Lee[8,9], Pushpendra Gupta[3], Liam G.Connolloy[5,10], Tzu-Hsuan Hsu[11], Jack Kramer[11], Bryan T. Bosworth[1], Nicholas R. Jungwirth[1], Eric J. Marksz[1], Aaron Hagerstrom[1], Tomasz Karpisz[1,2], Arundhati Ghosal[12], Lane W. Martin[7,8,9,13], Yimo Han[7,8,9,14], Angela C. Stelson[1], Christian J. Long[1], Ruochen Lu[11], Lucas Caretta[12,15], Javier Junquera[16], Jason J. Gorman[5], Long-Qing Chen[4], Ramamoorthy Ramesh[3], and Nathan D. Orloff[1]

* These authors contributed equally

[1] Communications Technology Laboratory, National Institute of Standards and Technology, Boulder, CO, USA

[2] Department of Physics, University of Colorado Boulder, Boulder, CO, USA

[3] Department of Materials Science and Engineering, University of California, Berkeley, CA, USA

[4] Department of Materials Science and Engineering, Pennsylvania State University, University Park, PA, USA

[5]Microsystems and Nanotechnology Division, National Institute of Standards and Technology, Gaithersburg, MD, 20899, USA

[6] Physique Théorique des Matériaux, Q-MAT, Université de Liège, B-4000 Sart-Tilman, Belgium

[7] Department of Chemistry, Rice University, Houston, TX, USA

[8] Department of Materials Science and NanoEngineering, Rice University, Houston, TX, USA

[9] Rice Advanced Materials Institute, Rice University, Houston, TX, USA

[10]Whiting School of Engineering, Johns Hopkins University, Baltimore, MD, 21218, USA

[11] Department of Electrical and Computer Engineering, University of Texas, Austin, TX, USA

[12]School of Engineering, Brown University, Providence, RI, USA

[13]Department of Physics and Astronomy, Rice University, Houston, TX, USA

[14]Smalley-Curl Institute, Rice University, Houston, TX, USA

[15]Department of Physics, Brown University, Providence, RI, USA

[16]Departamento de Ciencias de la Tierra y Física de la Materia Condensada, Universidad de Cantabria, Cantabria, Spain

# Abstract

Resonators are a key component in modern communications and computing. As demand and technological advances push component requirements into the terahertz regime, there is significant research devoted to the search for resonances at these frequencies. While uniform solid-state materials usually do not intrinsically feature resonances in this frequency range, self-assembled periodic arrays of ferroelectric nanodomains may provide an engineering route to design millimeter-wave properties. Here, we utilize prototypical dielectric-ferroelectric $SrTiO_3$/$PbTiO_3$ superlattices to robustly design periodic ferroelectric nano-scale domains. Phase field simulations predict an emergent domain breathing mode in complex polar textures and state-of-the-art millimeter-wave characterization shows evidence for such emergent resonances up to hundreds of GHz.

Complex polar textures in these superlattices lead to emergent piezoelectric properties that also result in millimeter-wave resonances, which are predicted by second principles methods and confirmed by direct measurement. The principles investigated in this work suggest a new modality for ferroelectrics in the design of millimeter-wave electronics.

# Main Manuscript

## Introduction

High-frequency resonators are a critical part of modern electronics. Everything from modern cellphones, which have hundreds of resonators in each handset, to quantum computers, which rely on resonators for readout, buses, and isolation, are dependent on this technology. Below 10 GHz, acoustic resonators are an established part of everyday electronics, but the small wavelength of sound means that acoustic resonators at higher frequencies require devices on the order of single nanometers, beyond the limits of practical fabrication[1–3]. Like their acoustic counterparts, photonic resonators[4] use physical boundary conditions on the geometry to contain electromagnetic waves, which is limited by the device footprint. Despite decades of research, challenges in this terahertz gap, the range from 100-1000 GHz, remain a fundamental obstacle for both scientists and engineers. Electronics at these frequencies could enable terabit data rates and new sensing capabilities[5], however, finding physics that enables resonators not limited by geometry and fabrication is a daunting research challenge.

One hope is intrinsic material resonances at the relevant frequencies. Today, the only option is ferromagnetic resonance (FMR), however, the required external magnetic biasing[6,7] (several Tesla above 100 GHz) renders on-chip integration extremely difficult. Spin-wave devices also require biasing but additionally rely on physical boundary conditions[8,9]. Because of these limitations, a nonmagnetic system is clearly preferable, but there is no existing dielectric analogue to FMR. The open question is then: is it possible to engineer a device-size-independent resonance in an alternative materials platform? We hypothesize that precise control of ferroelectric textures can enable a bottom-up approach to creating material-intrinsic dielectric resonances.

Intentionally designed metamaterials, systems where an artificial lattice creates emergent resonant properties using the periodicity of the structure[10,11], are well known in the acoustics, optics, and magnetics communities[12–14]. If ferroelectric domain structures could be precisely and reliably engineered, highly monodisperse, periodic arrays of ferroelectric nanodomains may feature metamaterial resonances in the millimeter wave-regime. First hypothesized by Pertsev and Arlt[15] in bulk ceramics at low frequencies (< 1 GHz), this concept of a domain lattice response could be extended to periodic array of ferroelectric domains on the scale of nm, resulting in resonances in the 100s of GHz.

Recent research on nanoscale domains in ferroelectric thin films[16–19] demonstrates the opportunity to engineer emergent physics. Artificial lattices of alternating ferroelectric and dielectric materials manifest tunable, but well-defined polar nanostructures[20–24], including emergent polar topologies. Such emergent polar vortices and skyrmions can additionally lead to other emergent phenomena[24–27] in the millimeter-wave regime, beyond what is associated with their periodicity. In these systems, the polarization state is robustly stabilized through a combination of elastic and electrostatic boundary conditions, commonly where the out-of-plane polar state is frustrated by the presence of an adjacent dielectric layer[23,28,29].

Here, we investigate well-defined, uniform nanoscale domains in the prototypical $(PbTiO_3)_n/(SrTiO_3)_n$ superlattices (PTO/STO, $n$ = 6, 12, 16, 20, 24, and 36 unit cells)[21,29]. In these superlattices, different epitaxial periodicities ($n$) give rise to unique polarizations textures including a fully in-plane polarized ferroelectric $a_1/a_2$ phase for short periods, an emergent vortex phase for intermediate periods, and a ferroelectric flux-closure phase for long periods[29]. In general, however, the critical dimension of the superlattice layer thickness ($n$)[30], creating an avenue for engineering the nanoscale polar texture. Engineered domains in superlattices offer significant practical advantages over those in single-layer ferroelectrics. Since their domain texture is largely controlled by electrostatic boundary conditions, such structures should be both much more robust to defects and of a repeatable and consistent size[31,32]. While the methodology proposed here is similar to that used in periodically poled LiNbO3 (PPLN) crystals, where ferroelectric domain walls are written as a way to control the cavity spacing, they are limited by the sizes of domains that can be achieved[33,34] (~1 um). Additionally, superlattice structures open a future avenue for 3D metacrystals of ordered 1D or 2D building blocks.

Existing work on ferroelectric superlattices has focused on simulation[35] , low-frequency characterization[25,36] and near-optical synchrotron-based studies to elucidate their physics[22,37,38]. Notably, however, high-frequency communications and computing applications rely on characterizing electrical properties at millimeter-wave frequencies, which is not directly probed by such techniques. Using cutting-edge, on-wafer dielectric spectroscopy, we can directly investigate the dielectric responses in the 100 GHz regime. Through phase-field simulations and direct electrical measurements, we show that by engineering the ferroelectric domain size and texture, we can intentionally create and observe dielectric resonances in the 100 GHz regime. Additionally, we observe emergent cross-coupled piezoelectricity that is associated with the complex ferroelectric topology, creating a functionality beyond what is set by the nanodomain periodicity. Enabled by state-of-the-art millimeter-wave measurements, we present a new modality establishing ferroelectrics as a platform for geometry-independent resonators in millimeter-wave electronics.

## Predicting resonances in nanoscale domains

In ferroelectric materials, ferroelastic domain walls act as electro-acoustic transducers. If the frequency of the electromagnetic excitation gives rise to an acoustic wave with the wavelength twice the domain size $d$, the individual acoustic waves emitted interfere constructively. In highly monodisperse domains, this interference can lead to a macroscopic resonance with resonance frequency $f_{\mathrm{res}}$. In the simplest case, assuming only acoustic interactions and a domain-size-independent speed of sound, the resonance frequency follows $f_{\mathrm{res}} \sim 1/d$ (**Supplementary Fig. 1**). In this formalism, when considering typical sound velocities in a solid, domain sizes in the nanometer range would then result in resonances in the millimeter wave and sub-THz range.

In PTO/STO superlattices where the layer thickness is small (< 10-12 unit cells), the ferroelectric layer typically self-assembles into fully in-plane polarized $a_1/a_2$ domains in order to minimize electrostatic and elastic energy[29]. In such a system, one could expect that the piezoelectric resonance would follow this $1/d$ relation with domain size. Consistent with this expectation, phase-field simulations (**Methods** and **Supplementary Note 1**) show a clear resonance that scales with the periodicity of the domains (**Fig. 1a**). In such simulations, the domain-wall position follows the electric field (**Fig. 1b**), resulting in a sinusoidal displacement (**Fig. 1c**), the so-called breathing mode[39] in which domain walls oscillate symmetrically about their equilibrium positions. When the wavelength is twice the domain size, the induced polarization associated with the individual domain-wall displacements interferes constructively to form a dielectric resonance (**Fig. 1d**). As expected, the resonance frequency increases with decreasing domain size (**Fig. 1e** and **Supplementary Fig. 2**). The fact that the scaling of the resonance frequency does not exactly follow the $f_{\mathrm{res}} \sim 1/d$ relation[15] expected from a simple acoustic interaction suggests that additional electrostatic interactions become relevant at nanometer-scale domain sizes. Besides the $a_1/a_2$ domain breathing mode, phase-field simulations also predict a resonant excitation in the polar-vortex phase, referred to as the vortexon[40] (**Supplementary Fig. 3**). This prediction suggests that engineering a highly monodisperse structure of ferroelectric domain walls can be used to obtain material-intrinsic resonances at millimeter wave frequencies.

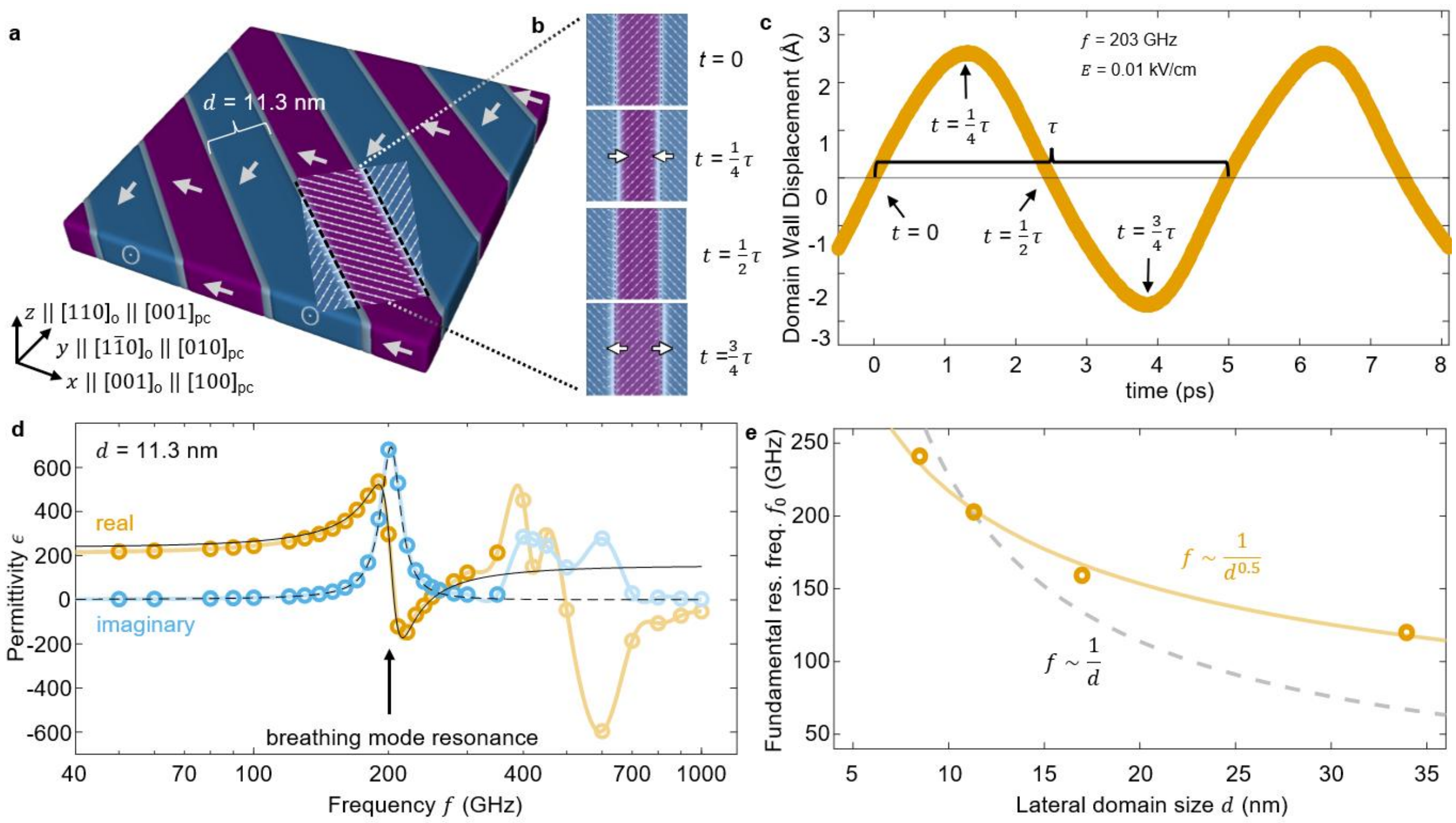

**Figure 1 | Phase-field simulations of resonances in $a_1/a_2$ domains.**

**a** Phase-field simulation showing the $a_1/a_2$ domain structure in a $PbTiO_3/SrTiO_3$ superlattice. The arrows show the direction of polarization. **b** Time resolved snapshots of the polarization field illustrating the breathing mode at $t$ = 0 ps, when the domain is smallest, and when the domain is largest. **c** Time resolved position of the domain walls in a, b, illustrating the predicted harmonic behavior. **d** Frequency dependent dielectric response of the $a_1/a_2$ breathing mode. The dispersions at higher frequencies correspond to higher order breathing modes or rotational modes[39]. **e** Dependence of the resonant frequency on the domain periodicity. The black dashed line shows the simple $f \sim 1/d$ prediction, the solid orange line shows the actual $f \sim 1/d^{0.5}$ trend of the simulated data.

## Designing nanoscale domains

Motivated by these predictions of the scaling with domain size, we deposited symmetric repetitions of PTO/STO heterostructures (**Fig. 2a**) to engineer targeted ferroelectric textures and directly probe these resonances electrically. 100-nm-thick superlattices of $(PTO)_n/(STO)_n$ ($n$ = 6, 12, 16, 20, 24, and 36 unit cells) were deposited using pulsed-laser deposition on single crystalline $DyScO_3$ (DSO) (110) substrates. Consistent with prior reports, this results in distinct nanoscale polar textures depending on the periodicity: a fully in-plane polarized ferroelectric $a_1/a_2$ phase (for $n$ = 6, 12), an emergent vortex phase (for $n$ = 16, 20, and 24), and a ferroelectric flux-closure phase (for $n$ = 36)[20,21,23,30] (**Fig. 2b-d**). *In-situ* RHEED and *ex-situ* X-ray diffraction confirm the quality of the deposition and thickness of the layers (**Supplementary Fig. 4** and **Supplementary Fig. 5**, respectively). In the polar-vortex structure, anisotropy results in the long axis of the polar vortex generally aligning along the $[1\bar{1}0]_o$. We refer to measurements along this direction as axial ($y$ || $[010]_{pc}$) and, orthogonally, lateral ($x$ || $[100]_{pc}$).

The periodic nanoscale domains arrange themselves into macroscale superdomains that are then observable with piezoresponse force microscopy (PFM). The ferroelectric domains self-assemble into superdomains of the $a_1/a_2$ domain variants ($n$ = 6, **Fig. 2e**), transitioning to "viscous" domains that are characteristic of vortex and flux closure structures ($n$ = 20, 36, **Fig. 2f-g**).

Cross-sectional high-angle annular dark-field scanning transmission electron microscopy (HAADF-STEM) images on $n$ = 6, 20, and 36 superlattices are (**Fig. 2m-p** and wide field STEM in **Supplementary Fig. 7**) overlaid with dipole maps calculated from the displacement of the titanium cations relative to the lead-cation (**Fig. 2h-k**). Such analyses confirm the expected phases, and combined with the RSMs, allow us to track the lateral spacing of the polar textures as a function of periodicity. In this regime, we observe that the critical polar texture spacing scales approximately linearly with the layer periodicity $n$ (**Fig. 2q**), even though the specific polar texture evolves from $a_1/a_2$ domains to polar vortices to flux-closure domains across this periodicity range. While our measurements suggest an approximately linear relationship, the underlying scaling may deviate from linearity across a broader range[41,42]. Having confirmed the distinct phases and domain sizes in the synthesized PTO/STO superlattices, validating the prediction of material-intrinsic resonances requires on-wafer dielectric spectroscopy extending into the millimeter-wave regime.

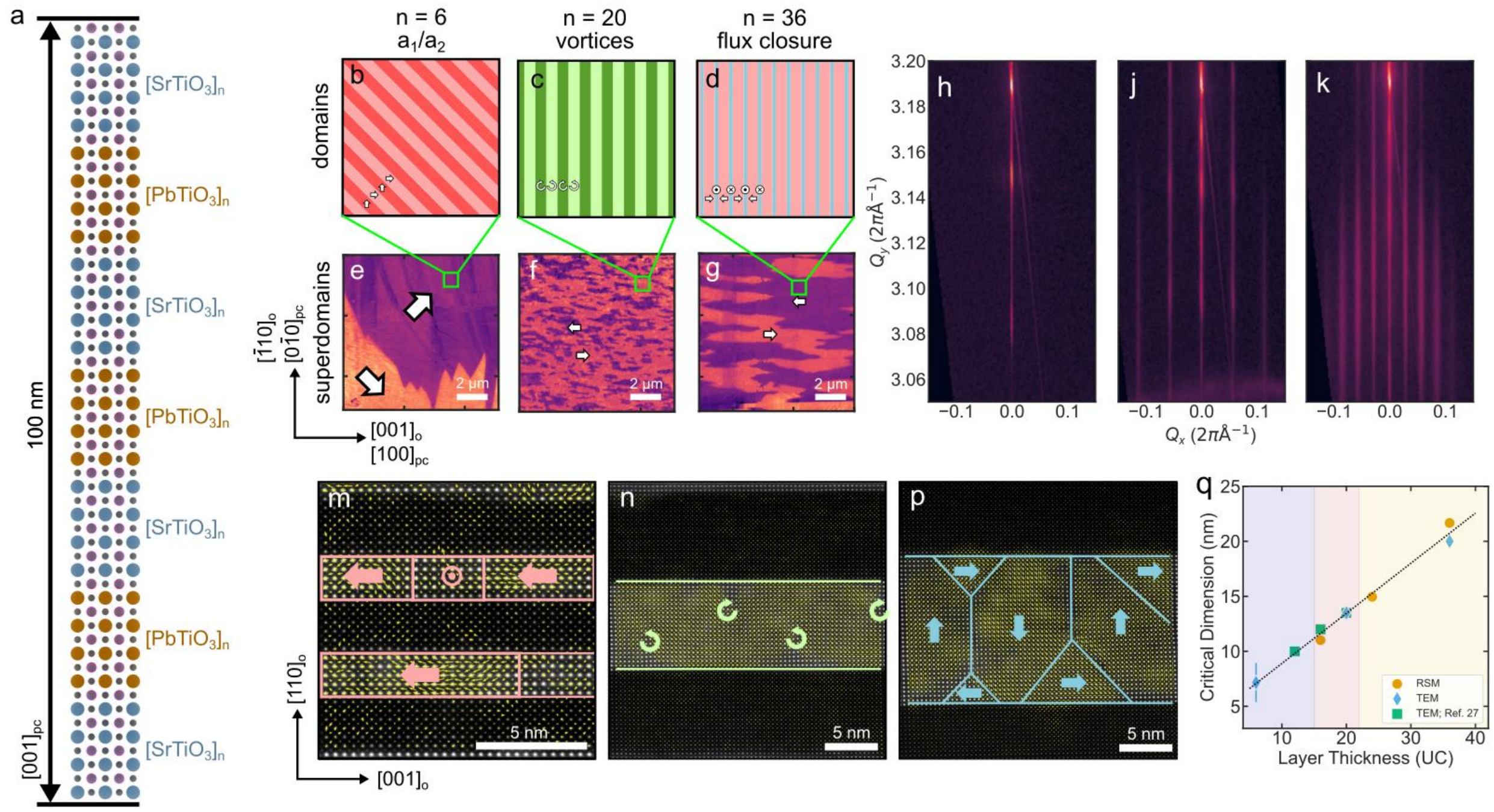

**Figure 2 | Designing domain structures from superlattice periodicity.**

**a** Illustration showing the superlattice structure made up of alternating layers of $PbTiO_3$ and $SrTiO_3$. **b-d** Illustrations showing the expected microscopic ferroelectric domain structure of $n$ = 6, 20, and 36 superlattices and, **e-g**, PFM of their corresponding superdomains. **h-k** RSMs of the structures, where the satellite peaks in the $Q_x$ direction are indicative of the domain spacing along the $[001]_o$ || $[100]_{pc}$ wavevector. **m-p** HAADF-STEM images, showing the atomic evolution of the texture from $a_1/a_2$ (m) to vortex (n) to flux closure (p) phase. **q** Scaling of the domains' critical dimension as a function of superlattice periodicity $n$, measured from both the RSMs and STEM data and in comparison to ref. [30]. Colored regions correspond to the expected a1/a2 (blue), vortex (red) and flux closure (yellow) phases.

## Observation of millimeter-wave resonances

As the wavelength of millimeter waves corresponds to features sizes in electrical circuits, material characterization at these frequencies requires an accurate, multi-tier calibration procedure (**Methods** and **Supplementary Note 2**). Such careful calibration is especially required for resolving spectral features such as narrow dielectric resonances. Our setup uses two perpendicular sets of coplanar waveguides (CPWs) oriented to measure the anisotropic complex in-plane permittivity[43] (**Fig. 3a**). The capacitance per unit length of these CPWs is sensitive to the in-plane permittivity of the thin film perpendicular to the propagation direction (**Fig. 3b**). Depending on the superlattice periodicity and the orientation, the films show distinct frequency dispersions (**Fig. 3c-e,** all films in **Supplementary Fig. 8**) that are indeed material-related, which we confirm by simultaneously fitting real and imaginary permittivity with a causal dispersion model. The sum of a Cole-Cole relaxation[44] and a harmonic oscillator resonance adequately describes all dispersions.

In superlattices hosting the $a_1/a_2$ phase ($n \leq 16$), we observe a resonance at frequencies around 100 GHz (**Fig. 3h,j**). The anisotropy in both resonance frequency and susceptibility of the resonating polarization mechanism is small, which is expected as the $a_1/a_2$ domain walls are oriented along the $[1\bar{1}0]_{pc}$, 45° to both sets of perpendicular CPWs. Despite its lower quality factor ($Q \approx 1$, **Supplementary Fig. 9**), the fact that the observed dispersion is narrower than a Debye relaxation constitutes clear evidence of a resonant, rather than relaxational, response (**Supplementary Fig. 10**). Additionally, this resonance frequency depends on the periodicity $n$ (**Fig. 3f**), indicating that it is the breathing-mode resonance predicted by our phase-field simulations. While broad relaxations are a known feature in ferroelectric thin films at millimeter-wave frequencies[45,46], a material-intrinsic dielectric *resonance* in this frequency range has not previously been reported.

Additionally, we observe resonances in superlattices hosting the vortex or flux-closure phases ($n \geq 16$). These resonances are only visible in the axial ($y$) direction and, in contrast to those in the $a_1/a_2$ phase, the resonance frequencies are independent of the periodicity $n$ (**Fig. 3g**). This independence from periodicity rules out the breathing-mode mechanism responsible for the $a_1/a_2$ resonances. Instead, the magnitude of the resonance frequency (≈ 25 GHz) suggests a piezoelectrically excited thickness shear mode of the film as the origin of the resonance, considering the speed of sound and the thickness of the film (**Supplementary Fig. 11**). Exciting such a thickness shear mode does not require an intrinsic material resonance but is instead driven geometrically by the film thickness, implying that the vortex and flux-closure textures exhibit unexpectedly large piezoelectric coefficients in their axial direction. Furthermore, the electric field-dependent resonance frequency resembles the typical butterfly shape of strain in a piezoelectric (**Supplementary Fig. 12**).

In the superlattices hosting the vortex or the flux-closure phases ($n \geq 16$) the dispersion is dominated by a broad relaxation (**Fig. 3h,j**), the center frequencies of which in the $n$ = 16, 20, and 24 superlattices are ($296.3 \pm 16.2$) GHz, ($145.6 \pm 1.5$) GHz, and ($55.2 \pm 0.7$) GHz, respectively. These frequency values suggest that this relaxation may be related to a collective dynamical mode of the vortex cores, referred to as the vortexon, which infrared spectroscopy has observed in this frequency range[40] (≈ 80 GHz) and is predicted by phase-field simulations (**Supplementary Fig. 3**). The anisotropic contribution to the permittivity in the orientation lateral to the vortex cores supports this interpretation (**Fig. 3j**). While the measurements confirm the predicted resonant character of the breathing mode of the $a_1/a_2$ phase, we do not see evidence of a resonant feature in the vortex phase in this frequency range. This contrasting result suggests that the phase-field simulations capture the nature of breathing mode accurately but there may be an unaccounted loss mechanism at higher frequencies. These measurements provide experimental benchmarks that can inform future simulations and guide the engineering of lower-loss resonances. Crucially, the on-wafer electrical metrology demonstrated here provides direct access to these emergent phonon modes without synchrotron infrastructure, opening a practical route to their study.

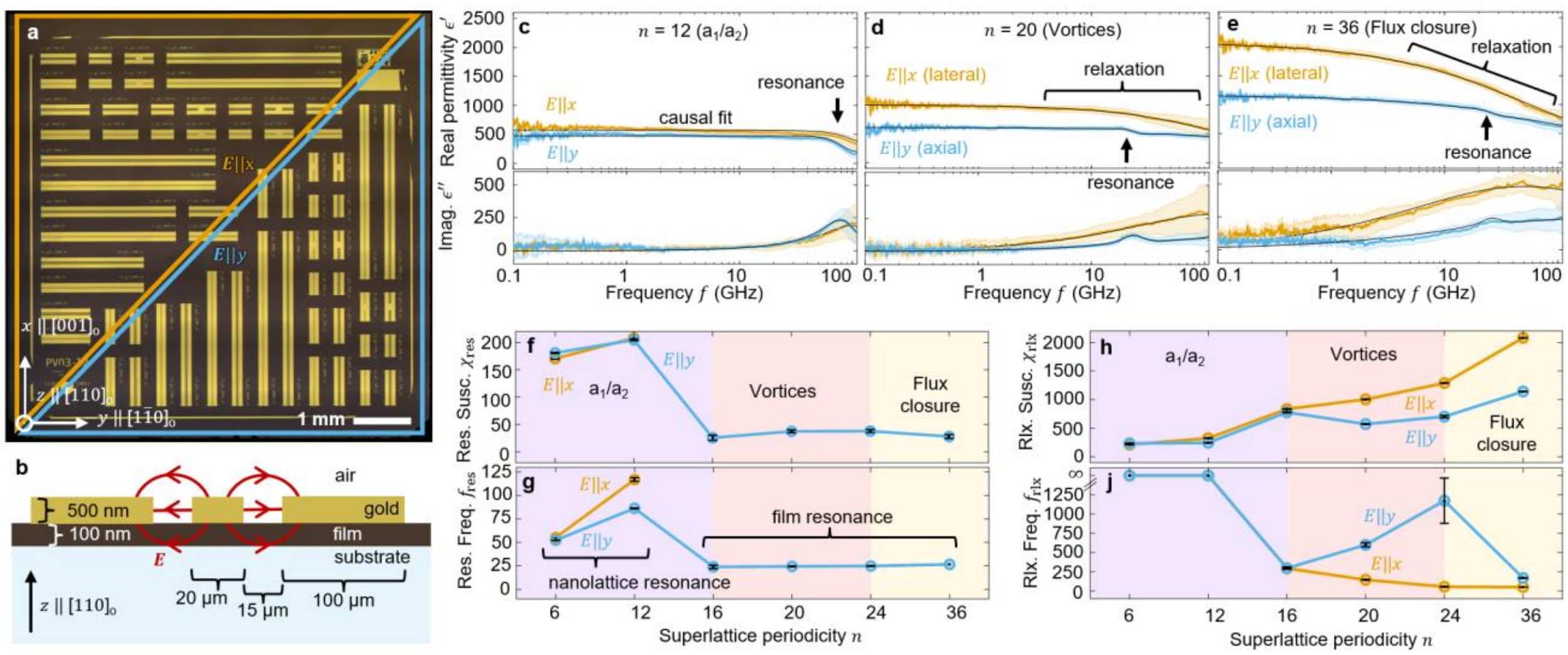

**Figure 3 | Observing resonances with on-wafer dielectric spectroscopy.**

**a** Photo of the CPW devices, differentiating between devices probing the lateral $(x)$ and axial $(y)$ directions on the sample. **b** Schematic of the CPW cross section with electric field distribution. **c-e** Frequency-dependent real (top) and imaginary (bottom) anisotropic permittivity of the $n$ = 12, 20, and 36 films in the $a_1/a_2$, vortex, and flux closure phases respectively. Black lines are causal fits and the shaded area the estimated standard uncertainty of the data. **f,g** resonance and **h,j** relaxation fit parameters for different superlattice periodicities, with standard fit uncertainties. The resonance frequency of $n$ = 36 in **f** and the relaxation frequencies for $n$ = 6 and $n$ = 12 in **j** do not have uncertainties we fix these frequencies to avoid fitting the measurement noise.

## Piezoelectric cross-coupling in polar vortices

Beyond the lattice modes that result from the monodisperse periodicity of the nanoscale-polar textures, these complex structures are expected to host other emergent electrical properties in the millimeter-wave regime[40]. As a specific example, the intertwined in-plane and out-of-plane polarization in the vortices (**Fig. 4a**) suggests strongly cross-coupled in- and out-of-plane material responses[47]. Both second-principles calculations (**Fig. 4b,c**) and phase-field simulations (**Fig. 4d**) predict substantial cross-coupled piezoelectric coefficients ($d_{133}^{\mathrm{eff}}$ = - 26 pm/V and - (72.8 $\pm$ 0.4) pm/V, respectively) arising from the vertical displacement of the polar-vortex cores due to an applied in-plane electric field. We use the convention $\varepsilon_{jk} = d_{ijk}E_i$ where $i, j, k = 1$ corresponds to the [100]pc, and $i, j, k = 3$ corresponds to the [001]pc or out-of-plane direction. In the quasi-steady-state, this coupling is related to the emergent texture of the phase[26] and is on the order of 5 to 10 times greater than that of PTO alone[48–50] (5 pm/V to 10 pm/V). Remarkably, this places the effective coefficient within the range of PZT[51] (50 pm/V to 150 pm/V).

To experimentally evaluate these emergent phenomena, we utilize state-of-the-art pulsed-laser interferometry[52] to directly detect out-of-plane deflection under an applied alternating field. The observed standing wave with wavelength matching the field along the transmission line confirms that the out-of-plane deflection is driven by the local electric field amplitude (**Fig. 4e,f**). When the electric field is applied along the [100]pc (lateral, $x$), orthogonal to the vortex axes, we observe a broad deflection peak around 30 GHz. While the frequency of this peak could stem from a compression mode (similar to the 25 GHz shear mode), the anisotropy suggests a genuine materials origin. The amplitude of this deflection corresponds to a piezoelectric coefficient $|d_{133}^{eff}|$ = 20 pm/V to 50 pm/V (**Fig. 4g**). When the electric field is applied along the orthogonal (axial, $y$) direction, however, we observe a smaller deflection ($|d_{233}^{eff}|$ = 5 pm/V to 12 pm/V) comparable to the broadband response. This anisotropy conforms to previous quasistatic and optical measurements, in which an electric field modulates the buckling of the vortex structure[25,36]. When a dc bias electric field is applied, the lateral amplitude in the broad peak decreases (**Fig. 4g**), suggesting that the increased amplitude is related to the vortex structure, which can be tuned with bias voltage. In the axial direction, this decrease with bias voltage is negligible.

The data consistently indicate a cross-coupling in the vortex structures driven by the buckling mode of the vortices[40]. Although every polar crystal class permits at least some transverse piezoelectric coefficients, the nanoscale confinement and heterogenous strain distributions present in these polar topologies dramatically amplify this transverse piezoelectric response, yielding an effective piezoelectric coefficient nearly an order of magnitude larger than bulk PTO. We hypothesize that the continuous rotation of polarization in the vortex topology provides a general design principle: a domain structure coupling in-plane and out-of-plane polarization should exhibit similarly enhanced transverse piezoelectric coefficients.

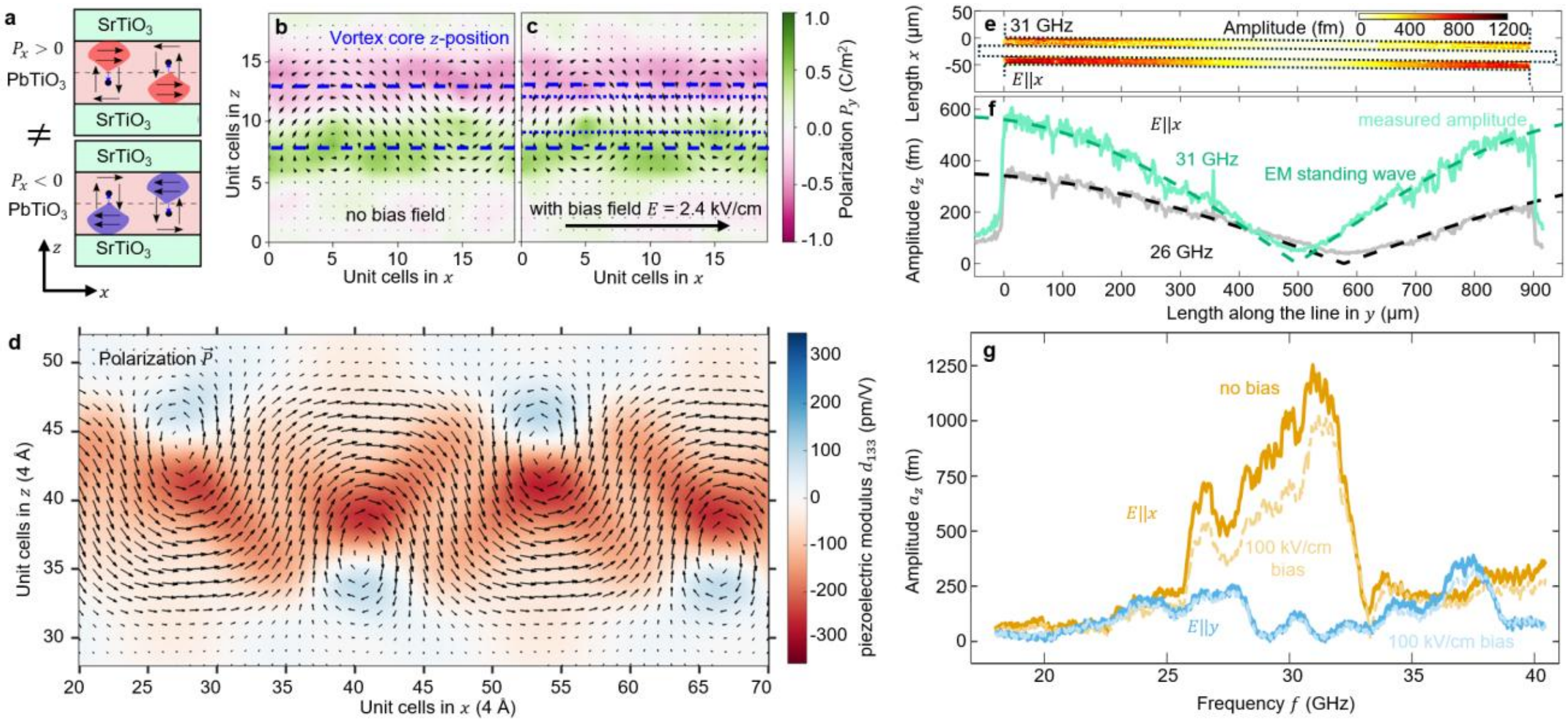

**Figure 4 | Prediction and measurement of emergent piezoelectricity.**

**a** Schematics of the off-center vortices breaking inversion symmetry. **b** Second-principles polarization maps showing vortex core position and **c** the displacement of the vortex cores under an electrical bias field. **d** Phase-field simulation showing the polarization map and the locally resolved piezoelectric modulus $d_{133}$. **e** Pulsed Laser Interferometry amplitude map at 31 GHz on the $n$ = 24 superlattice film. The dashed lines represent the edges of the Coplanar Waveguide (CPW) conductors. **f** Laterally ($x$) averaged displacement amplitude at 26 GHz and 31 GHz. The dashed lines represent the wavelength of the expected electromagnetic standing wave for each frequency, respectively. **g** Amplitude as a function of the electromagnetic signal frequency recorded at a single location at the end of the transmission line (see **Supplementary Fig. 15**).

## Conclusion

Although this work utilizes the well-studied PTO/STO superlattice system, our observation that engineered nanoscale ferroelectric polar textures act as a dielectric metamaterial establishes a design principle directly transferable to other piezoelectrically active ferroelectric systems. While ferroelectric films often show predictable equilibrium domain structures, ferroelectric-dielectric superlattices are expected to have significant advantages. The symmetric electrostatic discontinuity of the dielectric layer, compared to the free surface of a single ferroelectric film, should serve to create a much more stable and repeatable domain structure[53]. This kind of emergent domain structure has been shown in other ferroelectric-dielectric or ferroelectric-metal superlattice systems, which generalize the creation of complex, curling polarization textures[54–56].

Our phase-field simulations predict both the breathing mode in the $a_1/a_2$ phase and the vortexon in the vortex phase as resonant excitations. In contrast, our experiments showed that the breathing mode is indeed resonant but higher frequency modes appear to be overdamped. This discrepancy identifies a concrete target for future simulation of these systems: capturing the loss mechanisms that govern damping, with the goal of engineering low-loss topological resonance.

The precise engineering of boundary conditions in ferroelectric-dielectric superlattices enables the design of highly monodisperse ferroelectric domains. These domains act as an artificial lattice that can host emergent dielectric and piezoelectric properties. In this work, using state-of-the-art on-wafer dielectric spectroscopy, we directly show that this paradigm allows designing material-intrinsic dielectric resonances in the 100 GHz regime. By demonstrating that self-assembled ferroelectric texture, rather than device geometry, can be used to define resonant frequency, opening a materials-driven route toward closing the terahertz gap with potential consequences for on-chip millimeter-wave filters and quantum circuit elements alike.

# Methods

## Dynamical Phase-field simulations

We employ the dynamical phase-field method to simulate polarization and mechanical displacement dynamics that govern the frequency-dependent dielectric response. We also use it to simulate the piezoelectric response of the polar vortices. The evolution of the polarization is governed by the polarization dynamics equation

$$\mu_{ij}\frac{\partial^2 P_j}{\partial t^2} + \gamma_{ij}\frac{\partial P_j}{\partial t} = -\frac{\delta F}{\delta P_i}\ , \quad (1)$$

where $\mu_{ij}$ is the polarization effective mass, $\gamma_{ij}$ is the polarization damping constant and $F$ is the total free energy which is expressed as the integral over the following energy densities.

$$F = \int f_{Landau} + f_{Elastic} + f_{Electric} + f_{Gradient}\, dV\ . \quad (2)$$

For details on the individual contributions, see **Supplementary Note 1**. We define the superlattice structure by spatially varying the material properties (*i.e.,* dielectric stiffness coefficient) using a sharp interface description. The interface between the ferroelectric film and the $DyScO_3$ (DSO) substrate is treated as coherent allowing for the lattice mismatch strain between the DSO (110) surface and the ferroelectric layer to be calculated using the equivalent cubic lattice constant ( $a^c_{PbTiO3} = 3.9569$ Å , $a^c_{SrTiO3} =$ $3.905$ Å), and the pseudocubic lattice parameters of DSO $\left(0.5\sqrt{a^2+b^2}=3.946\text{ Å}, 0.5\,c=3.952\text{ Å}\right)$.[57] Since the lattices of the film and substrate are orthogonal there is no in-plane misfit shear strain such that:

$$\varepsilon_{11}=\frac{a_{DSO}^{[100]}-a^c}{a^c},\varepsilon_{22}=\frac{a_{DSO}^{[010]}-a^c}{a^c},\varepsilon_{12}=\varepsilon_{21}=0. \quad (3)$$

The out-of-plane components of stress $(\sigma_{33},\sigma_{23},\sigma_{32},\sigma_{13},\sigma_{31})$ are set to 0. For the $a_1/a_2$ simulations, a system size of $120\Delta x_1\times 120\Delta x_2\times 12\Delta x_3$ is used. The thickness of the $PbTiO_3$ and $SrTiO_3$ layers are set to $6\,\Delta x_3$. Periodic boundary conditions were applied in all directions. In all simulations $\Delta x_1=\Delta x_2=\Delta x_3=0.4\text{ nm}$. To initiate the simulation, a single *$a_1/a_2$* superdomain was artificially generated with a defined periodicity and evolved to reach a steady state condition. To simulate the dielectric response, a sinusoidal electric field with a frequency ranging from 10 GHz to 1000 GHz with an amplitude of 0.01 kV/cm is applied to the system. The real and imaginary dielectric constant is found by fitting the time-dependent polarization response to a sine and cosine function.

To simulate the piezoelectric response of the polar vortices, a system size of $128\Delta x_1\times 2\Delta x_2\times 96\Delta x_3$ is used. The thickness of the $PbTiO_3$ and $SrTiO_3$ layers are set to $16\,\Delta x_3$. Periodic boundary conditions were applied in all directions. In all simulations $\Delta x_1=\Delta x_2=\Delta x_3=0.4\text{ nm}$. To initiate the simulation, the polarization was set to random noise with $|P^{random}|=0.01\frac{\text{C}}{\text{m}^2}$, then the simulation was poled along the $[100]_{pc}$ with an electric field $E_1=100\text{ kV/cm}$. To simulate the piezoelectric response, a 10 GHz sinusoidal electric field amplitude of 10 kV/cm is applied to the system. In these simulations, the elasticity is solved using the equilibrium approximation. The piezoelectric coefficient is found by taking the derivative of the strain with respect to the electrical field as the electric field approaches 0 V/m (± 2,000 V/m).

## Second Principles Calculations

The second-principles simulations were performed using the same methodology presented in previous works[58,59] as implemented in the Scale-UP package[60]. The interatomic potentials and the approach to simulate the interface are the same as in Ref.[59]. We impose an epitaxial constraint assuming in-plane lattice constants of $a$ = $b$ = 3.911 Å, forming an angle of $\gamma$ = 90 °, while allowing the out-of-plane lattice parameter and the associated lattice angles to fully relax in order to minimize the stress. The in-plane mechanical condition corresponds to a small tensile epitaxial strain of +0.25 % with respect to the reference structure used in previous works (where $a$ = $b$ = 3.901 Å) similar to the one used on Ref [26].

The supercell dimensions correspond to 20×2×[$(PbTiO_3)_{10}/(SrTiO_3)_{10}$]. The $y$-direction is chosen to be 2 unit cells. This size is sufficiently large to allow octahedral rotations in the SrTiO3 layers, which can affect the dielectric response of the material. Additionally, limiting the structure to only 2 unit cells along the $y$-direction helps speed up the calculations while maintaining the essential physics of the system since perfectly periodic stripe domains are expected at low temperatures.

All the simulations performed in this study are conducted at $T$ = 0 K. We solved the models by running Monte Carlo simulations comprising 10,000 thermalization sweeps, followed by another 10,000 steps to compute average values of the out-of-plane strains and local polarization. Local polarizations are computed within a linear approximation of the product of the Born effective charge tensor times the atomic displacements from the reference structure positions divided by the volume of the 5-atoms unit cell centered on the A-sites of the perovskite.

## Film deposition and structural characterization

Samples were deposited using RHEED-assisted pulsed laser deposition. PTO and STO were deposited at 600 °C in a 100 mTorr $O_2$-background pressure using 10 Hz, ~1.7 J $cm^{-2}$ pulses from a KrF 248 nm excimer laser. Postdeposition, samples were annealed in 30 Torr O2 at the growth temperature for 15 min. Lab-scale X-ray diffraction was performed on a Panalytical diffractometer with a Cu K source. RSMs were taken about the symmetric $[220]_o$ diffraction peak along the $[001]_o$ axis to access the satellite peaks. Samples were imaged using lateral PFM in an asylum MFP-3D.

Cross-sectional samples for STEM imaging were prepared using an FEI Helios 660 focused ion beam (FIB). During the ion milling process, the acceleration voltages were gradually reduced from 30 to 5 kV and finally to 2 kV to minimize amorphous layers on the surface. HAADF-STEM images were captured using a 300 kV spherical aberration-corrected transmission electron microscope (FEI Titan Themis3 S/TEM) with a convergence angle of 25 mrad. The atomic column positions were determined by the multiple-ellipse fitting method in CalAtom software[61]. Local cation displacements were calculated relative to the centroid of the four nearest-neighbor cations from the opposite sublattice. The Ti displacement vector in each unit cell is opposite to the polarization direction of $PbTiO_3$. Prior to atomic-column position fitting, the STEM images were denoised using an average background subtraction filter (ABSF), which reduced random noise in the resulting displacement vector maps.

## Device fabrication

We fabricated Coplanar Waveguide (CPW) devices using standard photolithography on chips with superlattices $n$ = 6, 12, 16, 20, 24, 30, 36, as well as 100 nm PTO, 100 nm STO, and a bare DSO substrate control. These devices are 6 transmission lines of lengths 0.420 mm, 0.720 mm, 1.040 mm, 2.300 mm, 3.060 mm, 4.000 mm and an offset short-circuit reflect standard to form a complete mTRL calibration set[62]. The devices consist of nominal 500 nm thick gold on a 10 nm titanium adhesion layer deposited with electron-beam vapor deposition. The nominal cross section has a 20-μm-wide center conductor, 15-μm-wide gaps and 100 μm wide ground planes (**Fig. 3b**). The layout consists of two perpendicular CPW sets to access both the $[100]_{pc}$ and $[010]_{pc}$ directions of the films (**Fig. 3a**). For calibration purposes, we fabricated another set of CPW transmission lines with the same nominal dimensions on a separate 50.8 mm diameter $(LaAlO_3)_{0.3}(Sr_2TaAlO_6)_{0.7}$ (LSAT) wafer. This layout on this wafer additionally includes a series resistor fabricated from a PdAu alloy, which we use to set the reference impedance on the LSAT substrate[63,64].

## Millimeter-wave material characterization

We recorded complex scattering parameters of on-wafer coplanar waveguide devices with a vector network analyzer. We extracted the complex permittivity from 100 MHz to 110 GHz with no dc bias voltage and from 100 MHz to 50 GHz with a dc bias voltage of up to 100 kV/cm (150 V / 15 μm). To obtain the permittivity, we first performed a first-tier calibration on an LSAT wafer to set the reference impedance with the multiline Thru-Reflect-Line algorithm[62] in combination with the series resistor calibration[64] (**Supplementary Fig. 16**). We then used this reference impedance and a second-tier mTRL calibration on a bare DSO substrate to extract the frequency-dependent propagation constant $\gamma_{\mathrm{DSO}}$ and the characteristic impedance $Z_{\mathrm{DSO}}$[65]. We assume the DSO substrate to be loss- and dispersionless in the frequency range of interest[43], which allows us to obtain the inductance $L$ and Resistance $R$ per unit length on the DSO substrate via:

$$i\omega L_{\mathrm{DSO}} + R_{\mathrm{DSO}} = \frac{\gamma_{\mathrm{DSO}}^2}{i\omega C_{\mathrm{DSO}}}, \tag{4}$$

where we obtained $C_{\mathrm{DSO}}$ by averaging $\mathrm{imag}(\gamma_{\mathrm{DSO}}/Z_{\mathrm{DSO}})/\omega$ between 2 GHz and 10 GHz. Note that the resistance $R_{\mathrm{DSO}}$ differs significantly from that obtained with quasi-static finite element simulations[66]. Assuming that the films are non-magnetic, we take this resistance $R_{\mathrm{DSO}}$ and inductance $L_{\mathrm{DSO}}$ per unit length on the DSO substrate to calculate the capacitance $C_{\mathrm{film}}$ and conductance $G_{\mathrm{film}}$ per unit length of the film samples:

$$i\omega C_{\mathrm{film}} + G_{\mathrm{film}} = \frac{\gamma_{\mathrm{film}}^2}{i\omega L_{\mathrm{DSO}} + R_{\mathrm{DSO}}} \tag{5}$$

Next, we mapped the $C_{\mathrm{film}}$ and $G_{\mathrm{film}}$ to the complex permittivity of the thin film $\epsilon$ using quasi-static finite-element simulations (**Supplementary Fig. 17**). Finally, we fitted the extracted complex permittivity $\epsilon$ as a function of frequency as the sum of a Cole-Cole relaxation and, if applicable, a harmonic oscillator resonance:

$$\epsilon(f) = \chi_{\mathrm{rlx}} \frac{1}{1 + (i\, f/f_{\mathrm{rlx}})^{1-\alpha}} + \chi_{\mathrm{res}} \frac{f_{\mathrm{res}}^2}{(f_{\mathrm{res}}^2 - f^2 - i\, f\, f_{\mathrm{res}}/Q_{\mathrm{res}})} + \epsilon_{f\to\infty} \tag{6}$$

All fit parameters are positive reals. $\chi_{\mathrm{rlx}}$ is the dielectric susceptibility of the relaxation far below its mean relaxation frequency $f_{\mathrm{rlx}}$. The parameter $0 < \alpha < 1$ is a measure of the spread of the distribution of relaxation times around that mean frequency $f_{\mathrm{rlx}}$[44]. Refer to **Supplementary Note 2** for more details on the millimeter-wave characterization.

### Pulsed Laser Interferometry

Pulsed laser interferometry (PLI) is a homodyne interferometric technique for measuring femtometer-scale vibrations at microwave frequencies[52]. A femtosecond mode-locked laser creates a frequency comb with a tunable repetition rate $f_p$. When a device is driven at an excitation frequency $f_{ex}$, its mechanical vibration phase-modulates the laser pulses. Optical mixing between the mechanical vibration and the frequency comb results in a beat frequency $f_b = |f_{ex} - n\, f_p|$. As the repetition rate is tuneable, this beat frequency can be chosen almost arbitrarily and can thus be detected by a low-bandwidth detector with high signal-to-noise-ratio. We estimate the microwave power at the device to be between 20 dBm and 25 dBm, resulting in electric fields in the order of several kV/cm.

## Data Availability

The figure data are available at the NIST Public Data Repository[67], additional data are available from the corresponding authors upon reasonable request.

## Acknowledgements

The authors thank L. Enright and J. C. Booth, both with the National Institute of Standards and Technology (NIST), and A. Surampalli at Rice University for their critical feedback on this paper's manuscript. F.B. thanks Marek Paściak and Jiří Hlinka, both at the Institute of Physics of the Czech Academy of Sciences, for guiding scientific discussions and A. Osella at NIST for support in cleanroom. This paper is an official contribution of NIST, usage of commercial products herein is for information only; it does not imply recommendation or endorsement by NIST. This work was supported by the CHIPS

Metrology Program, part of CHIPS for America, National Institute of Standards and Technology, U.S. Department of Commerce. P.M., A.R., L.Q.C., L.W.M., and R.R. acknowledge support from the Army Research Office under the ETHOS MURI via cooperative agreement W911NF-21-2-0162. L.W.M., R.R., and Y.M. acknowledge additional support from the National Science Foundation under Grant DMR-2329111. A.R. acknowledges the support of the National Science Foundation Graduate Research Fellowship Program under Grant No. DGE1255832. The phase-field simulations in this work were conducted with the advanced computing resources provided by Texas A&M High Performance Research Computing, through allocation MAT250123 from the ACCESS program, which is supported by National Science Foundation Grants Nos. 2138259, 2138286, 2138307, 2137603 and 2138296. F.G.O. acknowledges financial support from MSCA-PF 101148906 funded by the European Union and the Fonds de la Recherche Scientifique (FNRS) through the grant FNRS-CR 1.B.227.25F. J.J. acknowledges financial support from Grant No.~PID2022-139776NB-C63 funded by MICIU/AEI/10.13039/501100011033 and by ERDF “A way of making Europe” by the European Union. T.J.L. acknowledges that this material is based upon work supported by the Air Force Office of Scientific Research under award number FA9550-24-1-0266. Any opinions, findings, and conclusions or recommendations expressed in this material are those of the author(s) and do not necessarily reflect the views of the United States Air Force. The development of the pulsed laser interferometry measurement method applied in this work was partially supported by the Deutsche Forschungsgemeinschaft (DFG, German Research Foundation) – project number 520954566. K.Y., X.L., and Y.H. acknowledge the support from Welch Foundation (C-2065 and X-C-0011), ACS PRF (67236-DNI10), and NSF (FUSE-2329111 and CMMI-2239545) and Sustainability Institute & RAMI at Rice University. X.L. acknowledges support from the Rice Advanced Materials Institute (RAMI) at Rice University as a RAMI Postdoctoral Fellow. Unless otherwise noted, NIST work was funded solely by the United States Government.

# Supplementary Information

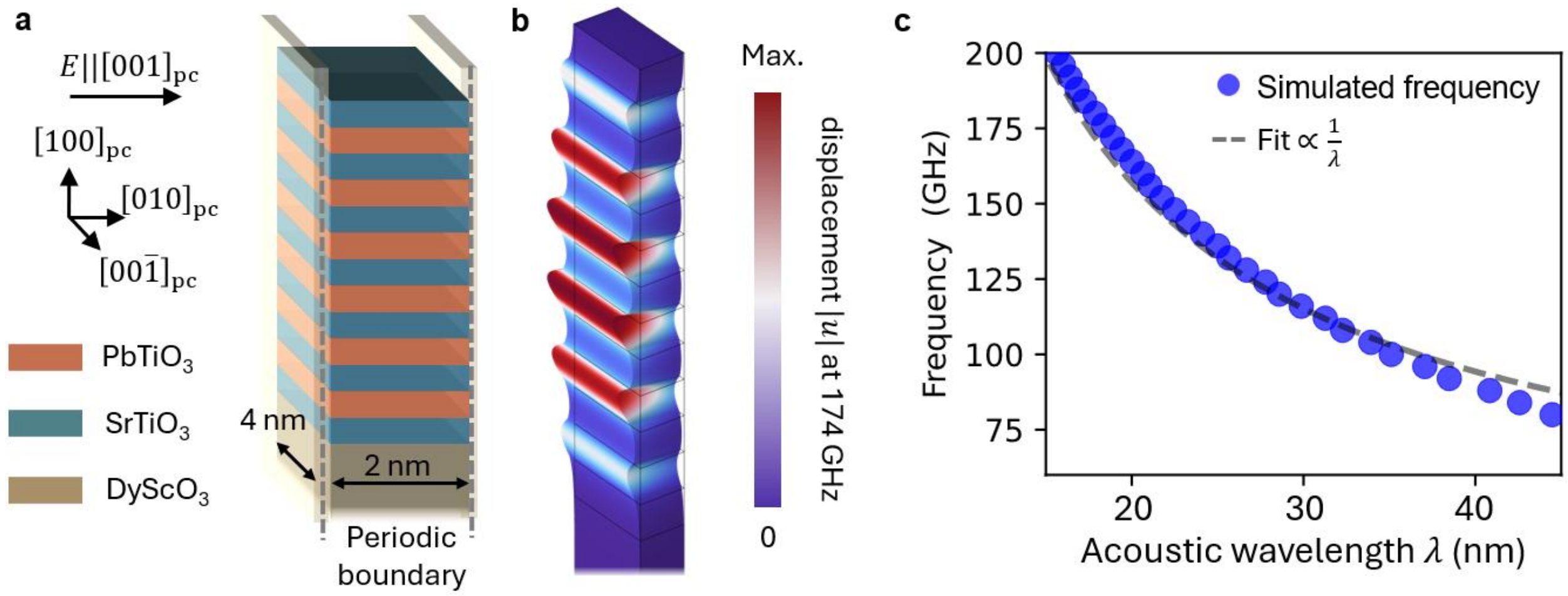

**Supplementary Figure 1 | Acoustic dispersion simulation.** Finite element acoustic simulation showing the dispersion relation expected from bulk material properties of PTO and STO. **a** Simulation setup showing the unit cell, **b** simulated displacement magnitude at 174 GHz. **c** Simulated admittance showing the expected dispersion relation $f \sim 1/\lambda$ that would lead to a resonance frequency dependence $f_{\text{res}} \sim 1/d$.

## Supplementary Note 1: Phase-Field Simulations

Following the Landau-Ginzburg-Devonshire Theory of Ferroelectricity[68], the Landau free-energy density describes the intrinsic stability of the ferroelectric phase with respect to the paraelectric phase and is written as a Taylor expansion of the polarization about the paraelectric phase (i.e. $P_i = 0$):

$$f_{Landau} = a_{11}(T)(P_1^2 + P_2^2 + P_3^2) + a_{1111}(P_1^4 + P_2^4 + P_3^4) + a_{1122}(P_1^2 P_2^2 + P_2^2 P_3^2 + P_1^2 P_3^2) + a_{111111}(P_1^6 + P_2^6 + P_3^6) + a_{111122}\left(P_1^4(P_2^2 + P_3^2) + P_2^4(P_1^2 + P_3^2) + P_3^4(P_1^2 + P_2^2)\right) + a_{112233} P_1^2 P_2^2 P_3^2,$$

where $a_{ij}$, $a_{ijkl}$, and $a_{ijklmn}$ are the dielectric stiffness coefficients measured under constant stress conditions. For $PbTiO_3$, a $6^{th}$-order expansion[69] is used and for $SrTiO_3$, a $4^{th}$ order expansion is used[70]. The dielectric stiffness coefficients are given in **Supplementary Tab. 1**.

The gradient energy density is

$$f_{gradient} = \frac{1}{2} G_{ijkl} \frac{\partial P_i}{\partial x_j} \frac{\partial P_k}{\partial x_l},$$

where $G_{ijkl}$ is the gradient energy tensor. The elastic energy density is

$$f_{elas} = \frac{1}{2} C_{ijkl} \left( \varepsilon_{ij} - \varepsilon_{ij}^0 \right) \left( \varepsilon_{kl} - \varepsilon_{kl}^0 \right),$$

where $C_{ijkl}$ is the elastic stiffness tensor, $\varepsilon_{ij}$ is the total strain (using the stress-free paraelectric phase as a reference), and $\varepsilon_{ij}^0$ is the eigenstrain which is related to the polarization by

$$\varepsilon_{ij}^0 = Q_{ijkl} P_k P_l,$$

where $Q_{ijkl}$ is the electrostrictive tensor. The elastic stiffness and electrostrictive coefficients are given in **Supplementary Tab. 1**. For simplicity, we assume the elastic stiffness coefficients are homogenous throughout the superlattice. To calculate the local strain, we use the mechanical displacement ($u_i$) where the evolution of the mechanical displacement is given by the electrodynamic equation

$$\rho \frac{\partial^2 u_i}{\partial t^2} = \nabla \cdot \left( \sigma_{ij} + \beta \frac{\partial \sigma_{ij}}{\partial t} \right),$$

where $\rho$ is the material mass density, $\beta$ is the stiffness damping coefficient and $\sigma_{ij}$ is the stress field.

The electrostatic energy density is

$$f_{elec} = -E_i P_i - \frac{1}{2} \epsilon_0 \kappa_{ij}^b E_i E_j,$$

where $\epsilon_0$ is the vacuum permittivity and $\kappa_{ij}^b$ is the background dielectric constant which is chosen as 20 and is isotropic. Where the inhomogeneous electrical field is solved assuming electrostatic equilibrium ($\nabla \kappa_{ij}^b \nabla \phi(x) = -\nabla \cdot P$) and the total electric field is equal to $E_i = -\nabla \phi + E_i^{ext}$, where $E_i^{ext}$ is the externally applied electric field. More details on numerical methods for the dynamical phase-field method can be found in the literature[71].

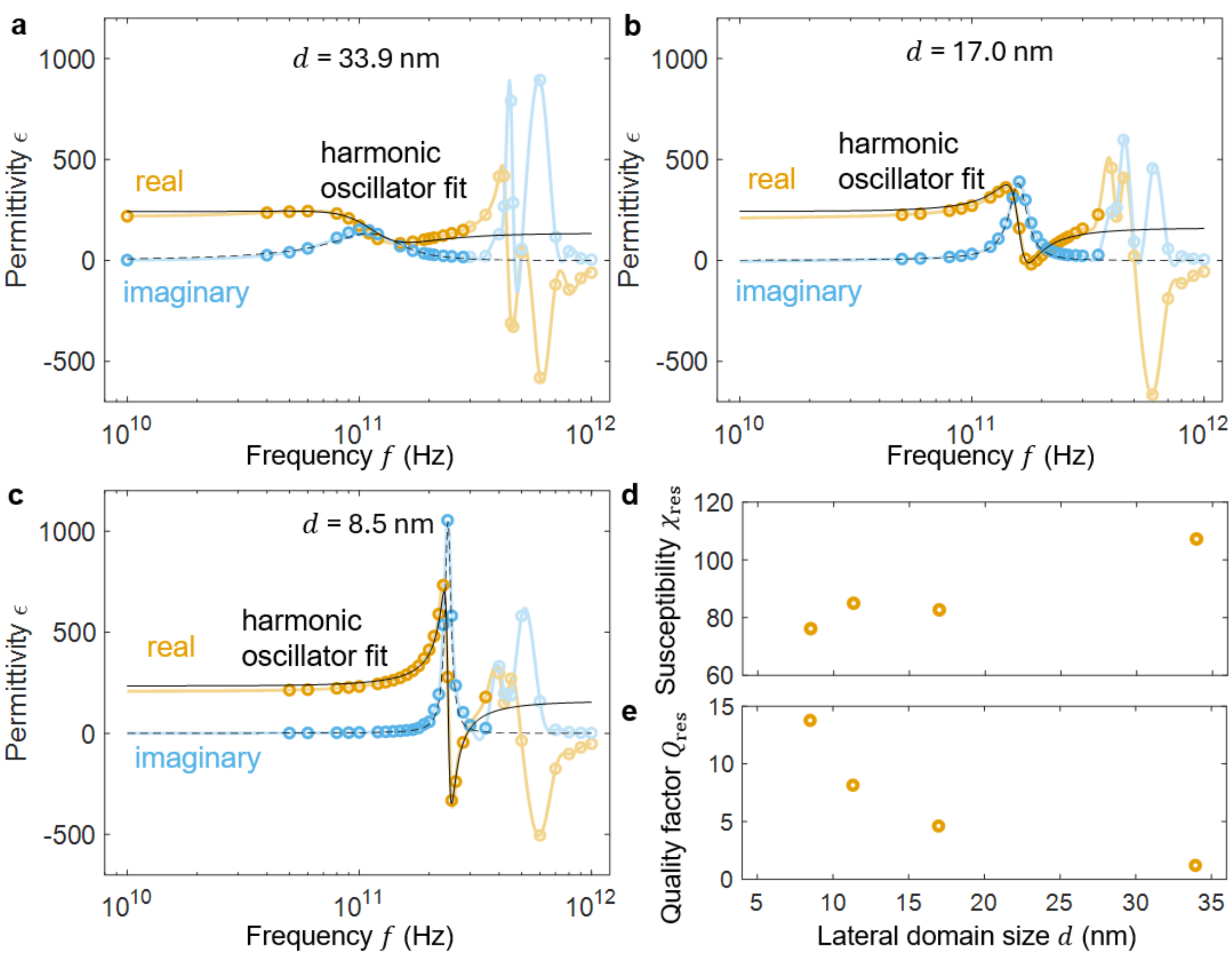

**Supplementary Figure 2 | Phase-field dispersions. a-c** Frequency dispersion of the complex permittivity for different lateral domain sizes using phase-field simulations. **d-e** Additional resonance fit parameters: susceptibility $\chi_{\mathrm{res}}$ of the resonance and quality factor $Q_{\mathrm{res}}$.

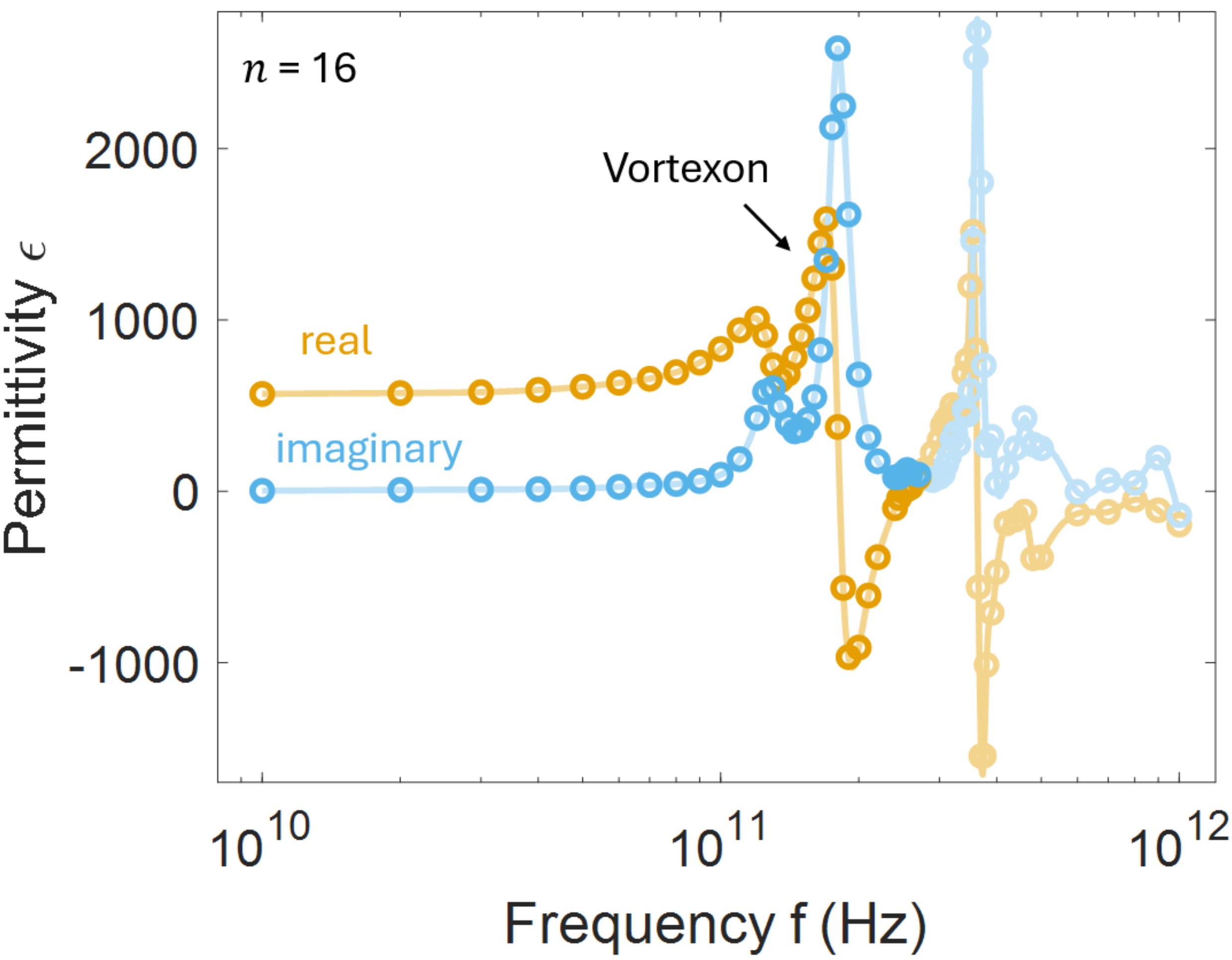

**Supplementary Figure 3 | Vortex phase frequency dispersion.** Frequency dispersion of the complex permittivity in the $n$ = 20 vortex phase obtained with Phase-field simulations.

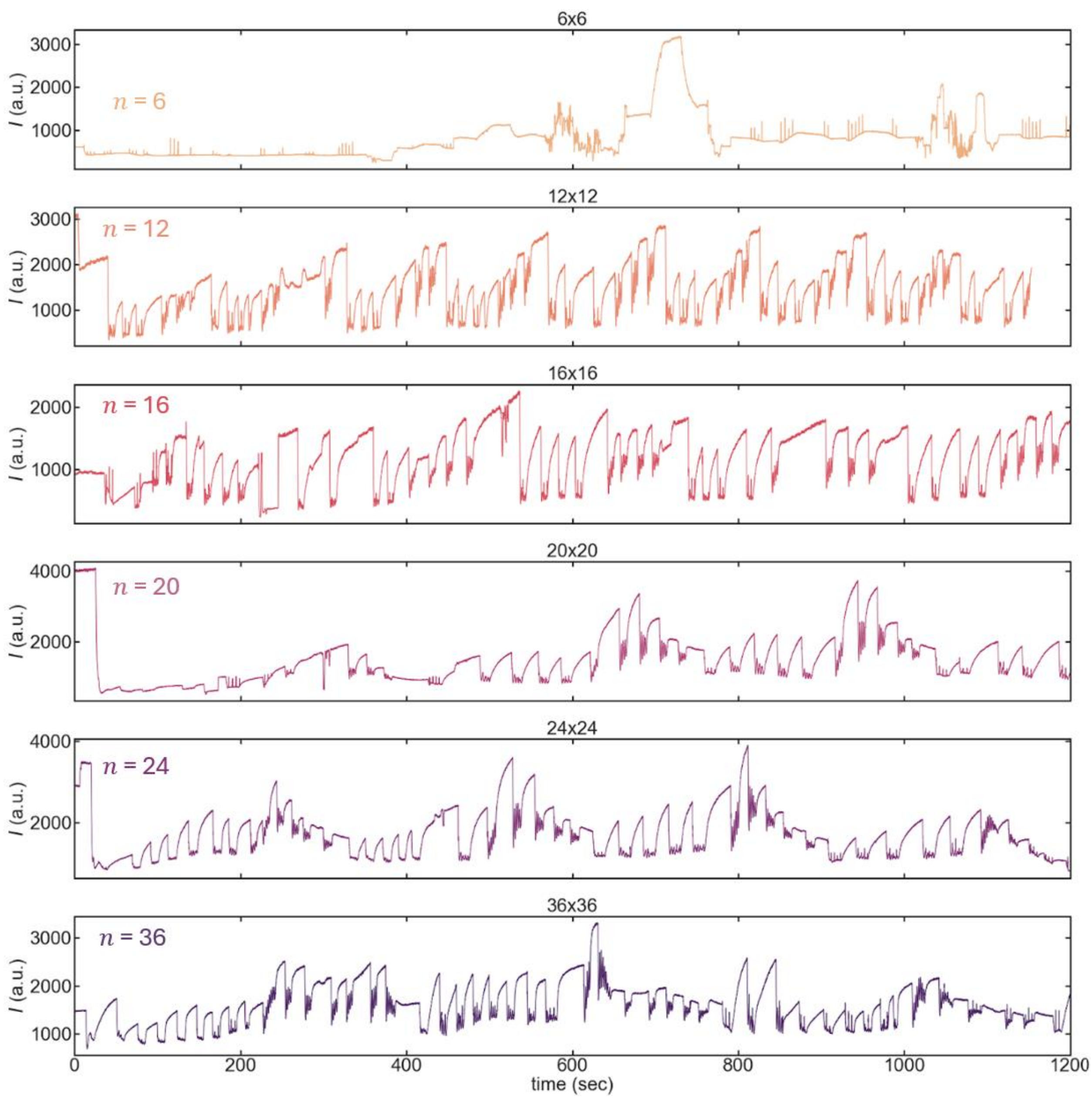

**Supplementary Figure 4 | RHEED oscillations.** RHEED spectra taken during sample deposition showing the layer-by-layer growth of PTO/STO superlattices. Layers are deposited in groups of 5 unit cells to allow for surface equilibration.

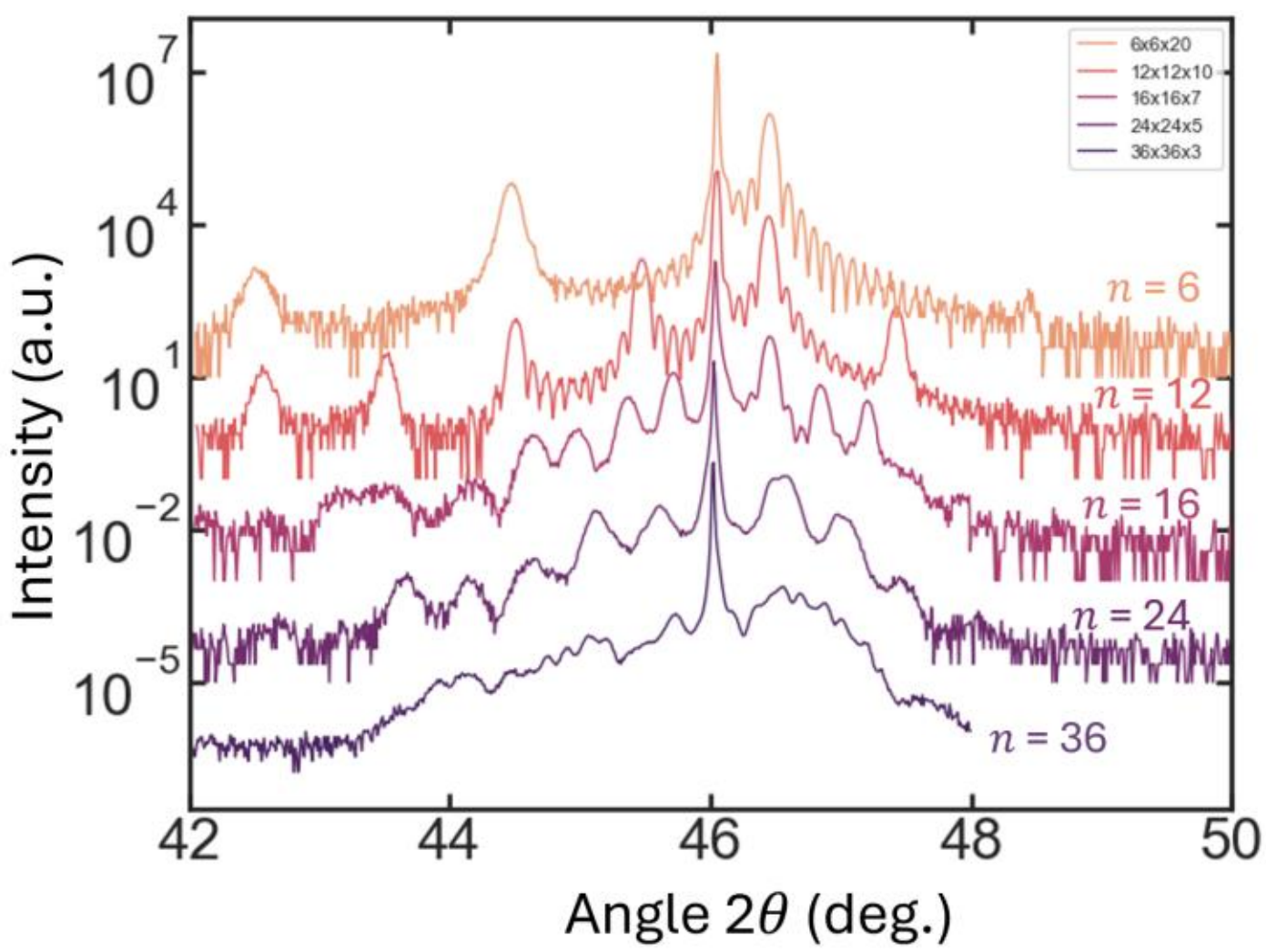

**Supplementary Figure 5 | 2$\theta$-$\omega$ X-ray diffraction.** Spectra of the $n$ = 6, 12, 16, 24, and 36 samples**.** In the thinner samples, the clear Laue fringes show excellent interfacial quality. In the samples where $n$ = 16 or 24, the phase splits into both $a_1/a_2$ and vortex contributions, each with their own effective out of plane lattice parameter, giving rise to two sets of superlattice reflections. In the $n$ = 36 sample, the superlattice structure is only repeated three times, making the superlattice peaks themselves hard to define.

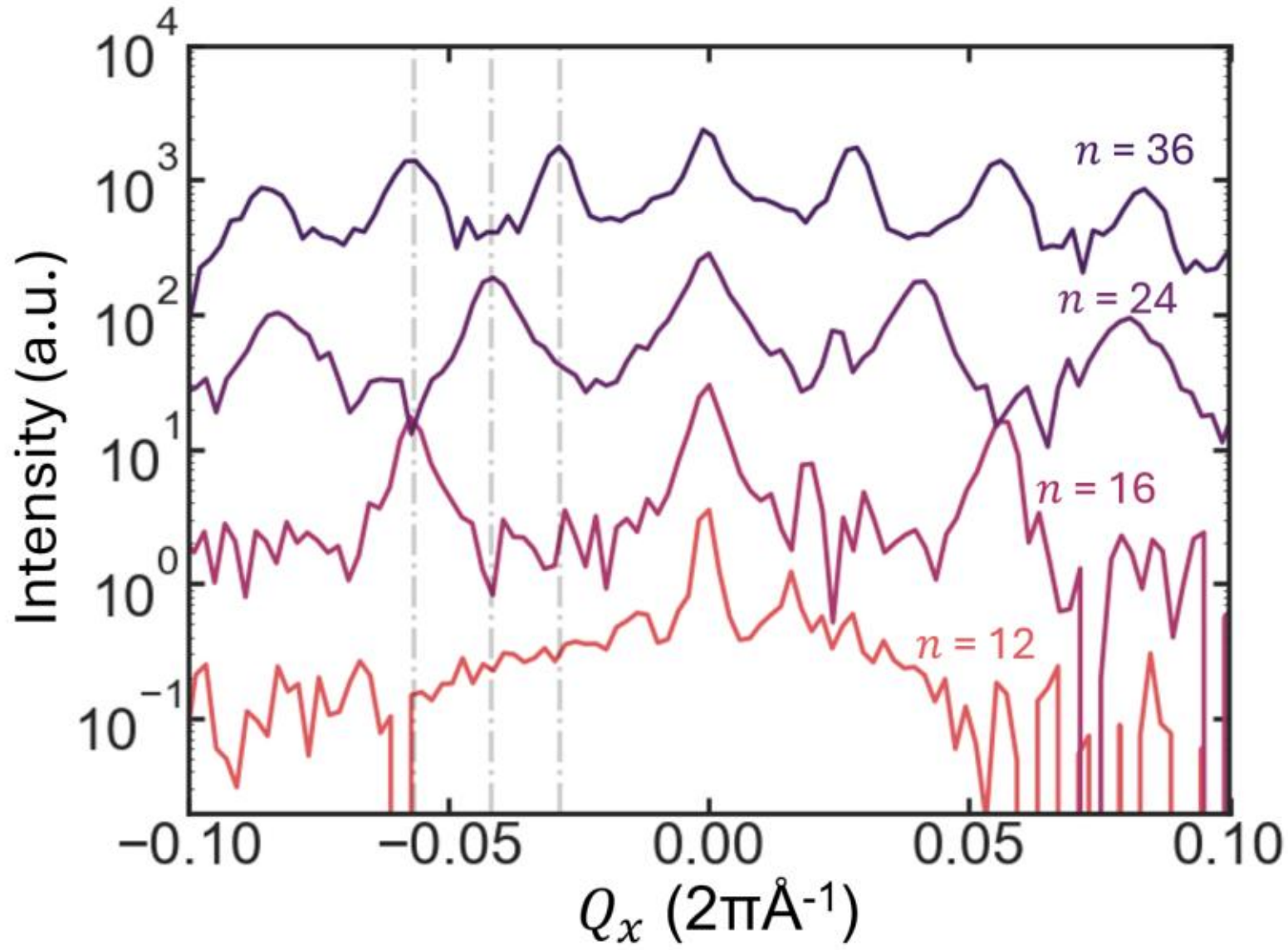

**Supplementary Figure 6 | In-plane periodicity.** $Q_x$ ($x$ || $[100]_{pc}$) cuts of RSMs shown in **Fig. 2h-k**, showing the in-plane periodicity of the structures. Colors correspond to those used in Supplementary Figure 4 representing, top to bottom, the $n$ = 36, 24, 16, and 12 samples. These data are extracted as horizontal line cuts from **Fig. 2m** at the first superlattice peak. The grey dotted lines show the position of the first peak.

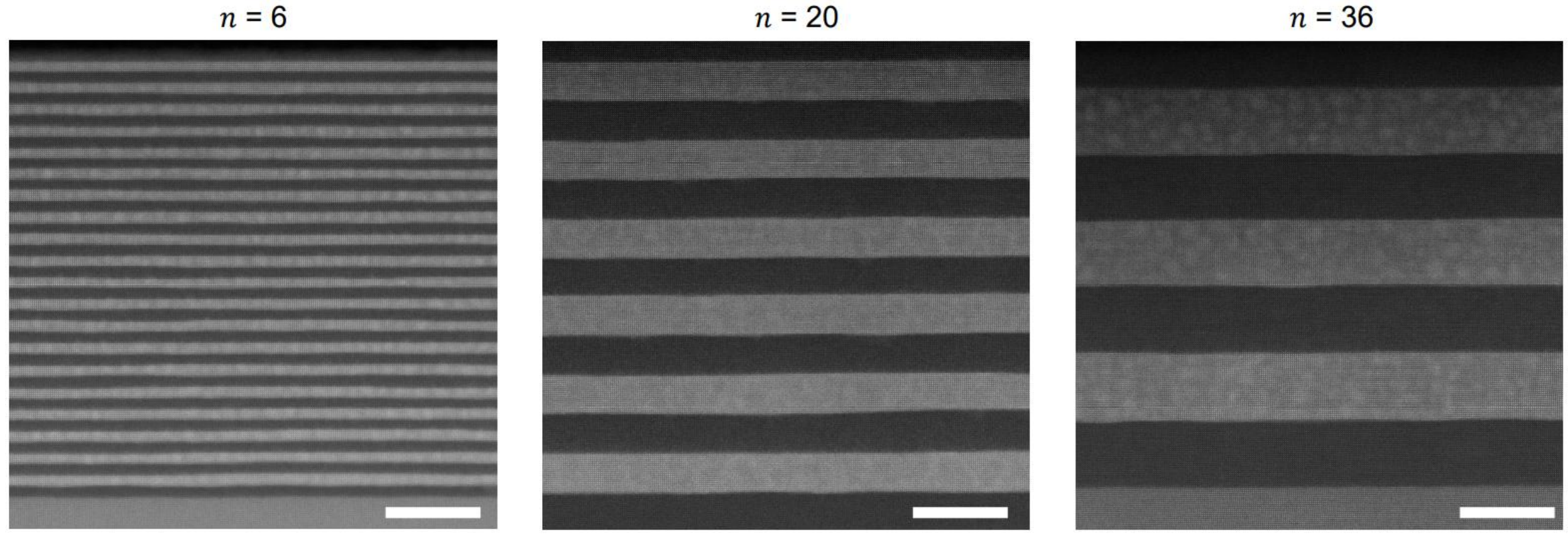

**Supplementary Figure 7 | Transmission electron microscopy.** Wide field of view HAADF-STEM images showing the n = 6, 20, and 36 PTO/STO superlattices. Scale bars are 20 nm.

## Supplementary Note 2: Millimeter-Wave Measurements

First, we performed a zero-dc bias measurement up to 110 GHz on the pristine films. Then, we cycled through a dc bias voltage of 150 V with a step size 7.5 V and measured the permittivity up to 50 GHz during that cycling (setup in **Supplementary Fig. 18**). Before measuring a full cycle of 0 V, ..., -150 V, ..., 150 V, ..., -150 V, ..., 0 V, we added an additional half cycle (0 V, ..., 150 V, ..., -150 V, ..., 0 V) to pole the film. We performed another zero-dc bias measurement up to 110 GHz after the cycling of the films. Note that the 25 GHz resonances in the $n$ = 20, 24, 36 films only became visible after poling, that is having cycled the film through $\pm$150 V (**Supplementary Fig. 8**).

The permittivity mapping assumes that the out-of-plane permittivity of the thin films does not exceed its in-plane permittivity by an order of magnitude. Then, the CPW geometry is only sensitive to the thin films in-plane permittivity that is perpendicular to the wave propagation direction in the CPW transmission line. As fabrication tolerance leads to chip-to-chip variation, we adjust the lateral dimension in the simulation according to the procedure described in Ref [72]. Not adjusting for this fabrication error can lead to noncausal frequency-dispersions of the complex permittivity.

We ruled out any magnetic properties of the films, specifically a magnetic origin of any resonance, by extracting all four circuit parameters without imposing a non-magnetic environment of the CPW transmission lines using the error boxes of the second-tier calibration[65]. This more general circuit fitting is noisier, but still sufficient to rule out magnetic film properties, specifically that the 25 GHz resonance is of magnetic origin (**Supplementary Fig. 19**).

Domain formation on the macroscale can make ferroelectric films highly inhomogeneous. This inhomogeneity across the chip poses a challenge to reliably extract film properties because the mTRL algorithm generally assumes a homogeneous propagation constant along and across all lines. The StatistiCal implementation[73] of the mTRL algorithm can accommodate for the inhomogeneity if there is sufficient phase accumulated. As we are measuring on a 10 mm x 10 mm chip, the longest line (4 mm) is not long enough to accumulate enough phase below 1 GHz (< 20 °). This small phase delay makes StatistiCal not be able to find appropriate weighting of the lines, which manifest in an apparent, non-causal dispersion of the extracted permittivity. Instead, we individually fitted the first-tier corrected S-parameters of the three longest lines by allowing $C_{\mathrm{film}}$ and $G_{\mathrm{film}}$ to vary while restricting the inductance and resistance to $R_{\mathrm{DSO}}$ and $L_{\mathrm{DSO}}$[74] and then use the fitted $C_{\mathrm{film}}$. In contrast, the accumulated loss is sufficient for StatistiCal and we therefore continue using the $G_{\mathrm{film}}$ extracted with StatistiCal.

To avoid overfitting when fitting the frequency dispersion of the complex permittivity, we set $\epsilon_{f\to\infty}$ = 1 in **Eq. 6** for most samples and only fitted a resonance when we visually observed a step in the dispersion of the real permittivity. Furthermore, the relaxation frequencies of the $n$ = 6 and $n$ = 12 samples were too far beyond the upper limit of our measurement range to reliably extrapolate it. For these two samples, we set $\chi_{\mathrm{rlx}}$ = 0 and allowed $\epsilon_{f\to\infty}$ to vary instead. For comparability, in **Fig. 3h,j** and **Supplementary Fig. 13a-c** we show this fitted $\epsilon_{f\to\infty}$ as a susceptibility $\chi_{\mathrm{rlx}} = \epsilon_{f\to\infty} - 1$ and $f_{\mathrm{rlx}} = \infty$.

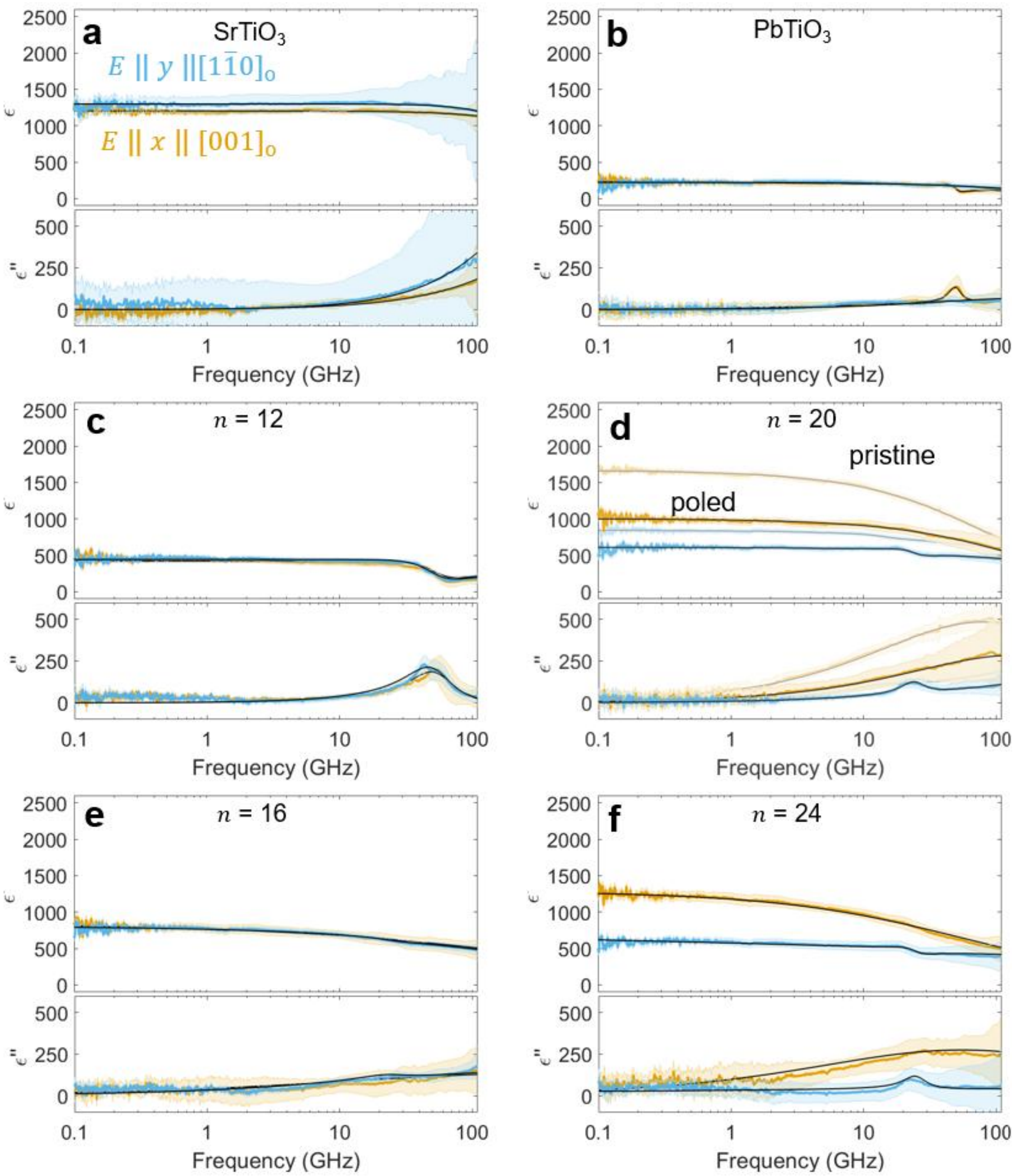

**Supplementary Figure 8 | Frequency dispersions.** Complex anisotropic permittivity ($\epsilon'$ real, $\epsilon''$ imaginary) of all measured films superlattices ($n$ = 6, 12, 16, 20, 24, 36) and two control films (STO and PTO). Observe the resonance in $PbTiO_3$ film in the $E$ || $[001]_o$ direction, which corresponds the domain breathing mode in PTO's domain structure. Interestingly, the resonance in the $n$ = 12 film is slightly split, which could indicate degeneracy in the macrodomains of the $a_1/a_2$ phase because the CPWs are sensitive to both macrodomains. Also observe how the resonance in the vortex film only emerges after poling.

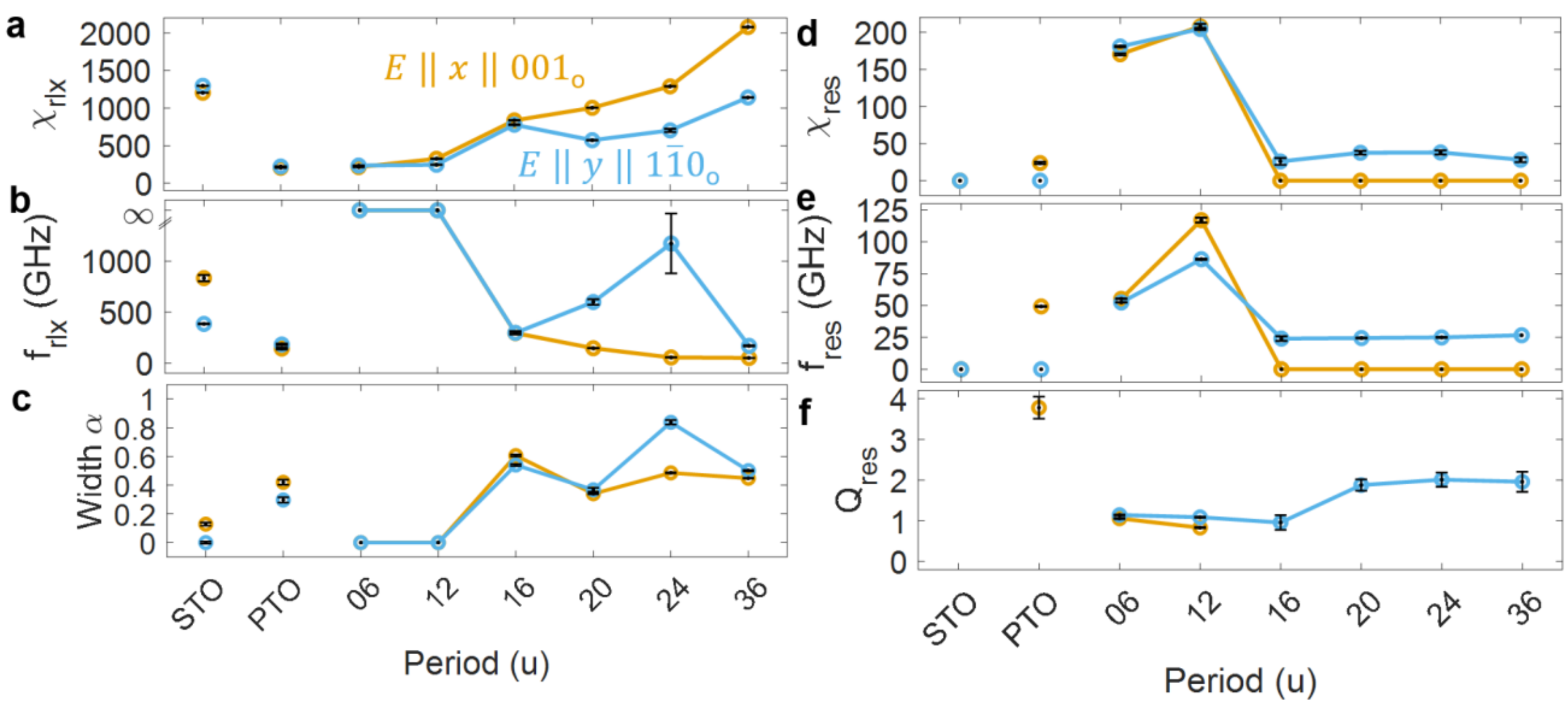

**Supplementary Figure 9 | Fit parameters of the frequency dispersions. a-c** fit parameters of the Cole-Cole relaxation (susceptibility $\chi_{\mathrm{rlx}}$, center frequency $f_{\mathrm{rlx}}$ and width $\alpha$). Note that for n = 6 and n = 12, the relaxation frequency is "infinite" because it is too far beyond the measured range to extrapolate. **d-f** fit parameters of the harmonic oscillator resonance (susceptibility $\chi_{\mathrm{res}}$, center frequency $f_{\mathrm{res}}$ and quality factor $Q_{\mathrm{res}}$).

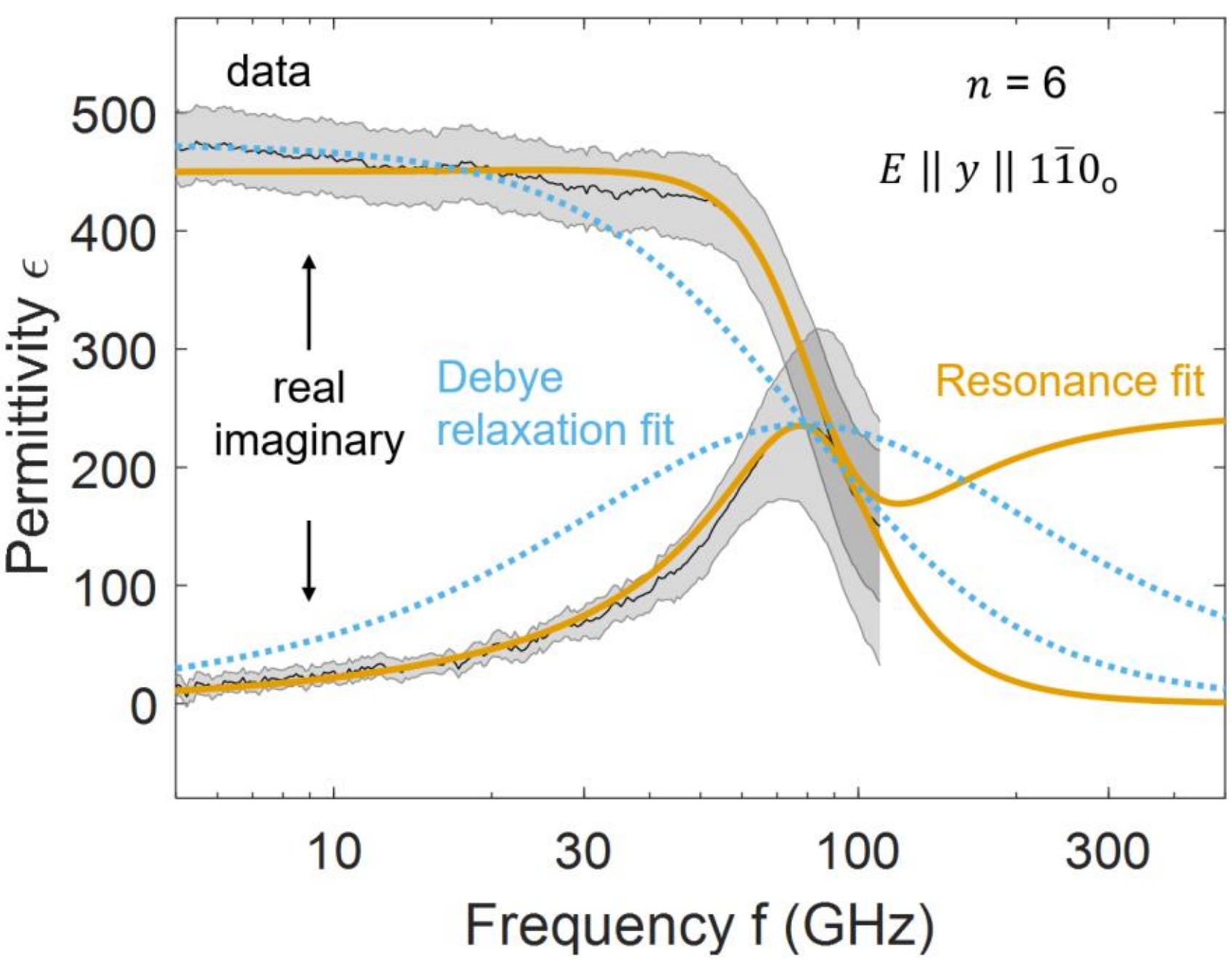

**Supplementary Figure 10 | Resonance *vs*. Relaxation.** Test showing that the observed dispersions are resonances not relaxations, at the example of the $n$ = 6 film in the $E$ || $[1\bar{1}0]_o$ || $[010]_{pc}$ direction.

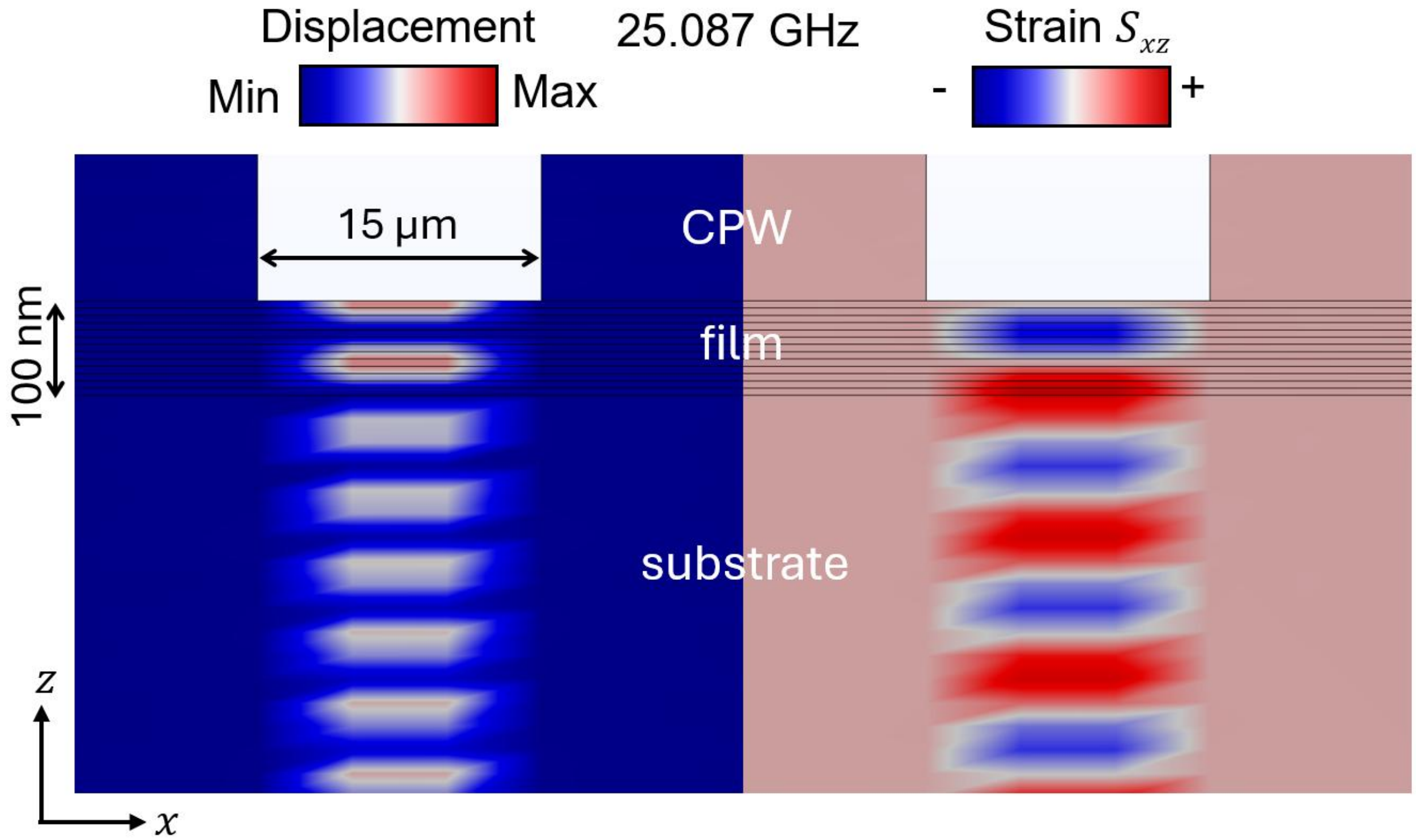

**Supplementary Figure 11 | Shear wave film resonance.** Finite element simulation of a 100 nm thick ferroelectric film on DSO substrate. The Eigenmode solver returns a resonance at 25 GHz.

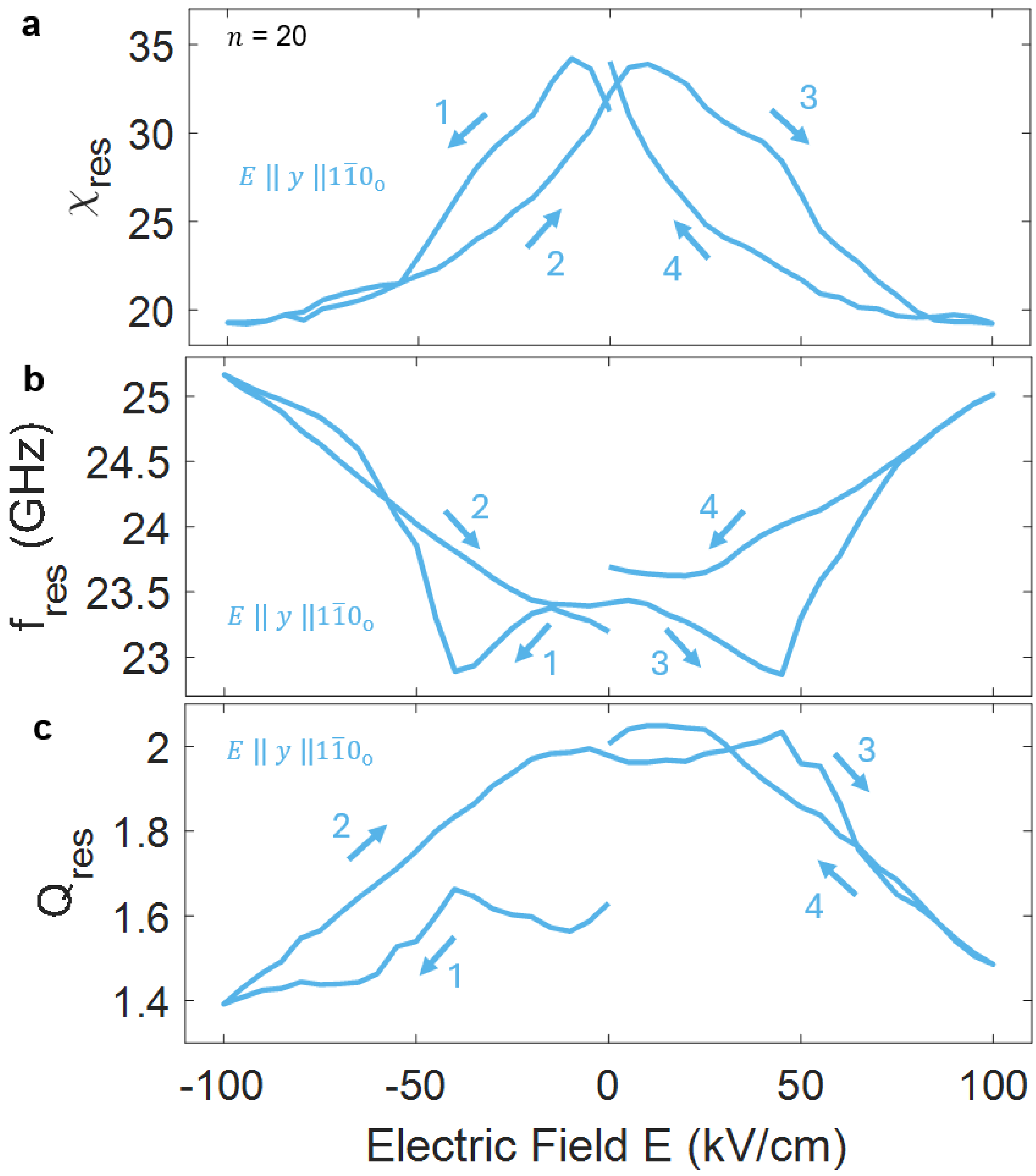

**Supplementary Figure 12 | Electric-field dependent resonance fit parameters.** Based on the sweeps up to 50 GHz. The resonance frequency $f_{res}$ in the $n$ = 20 film resembles the butterfly shape typical for strain-electric field curves in piezoelectrics.

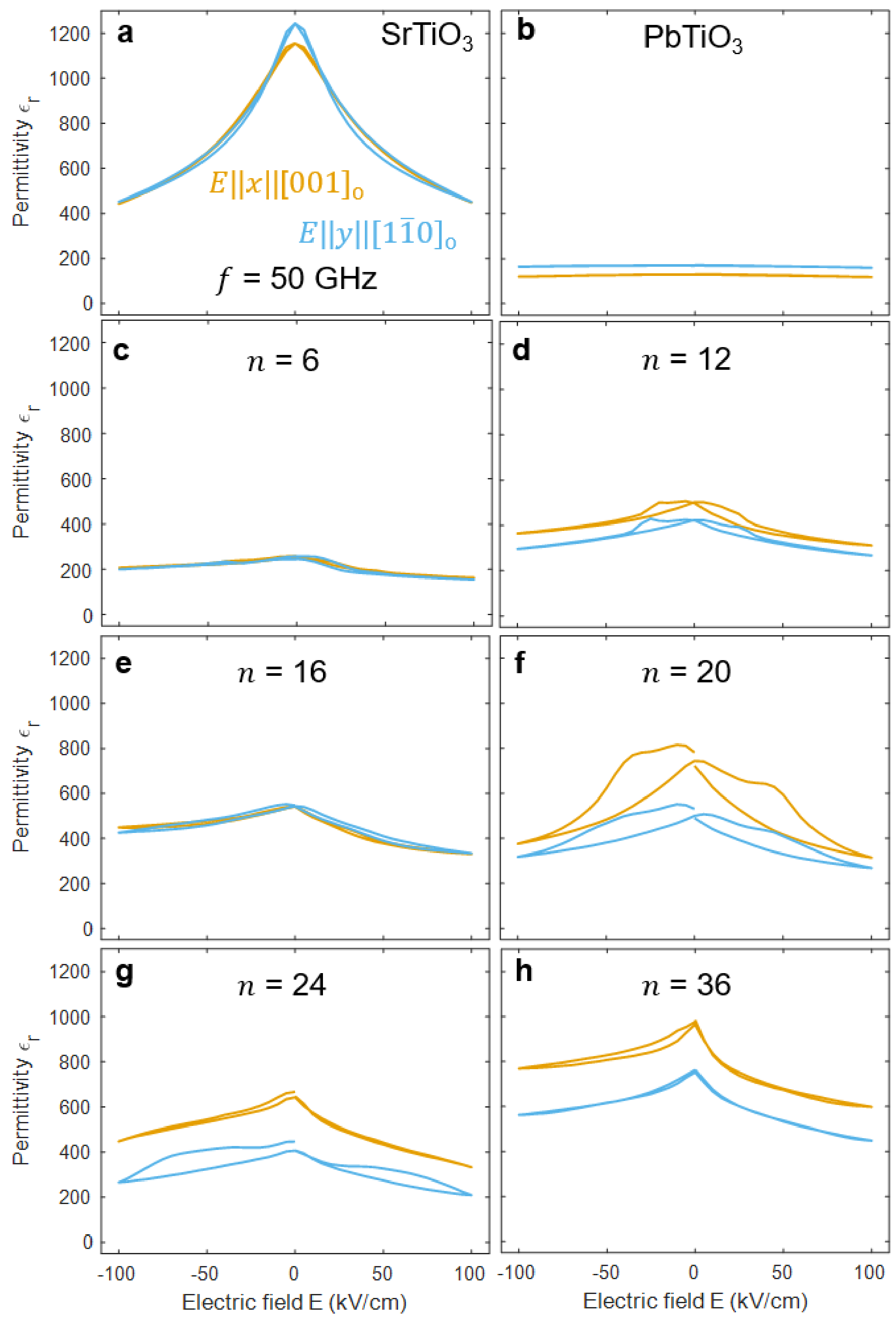

**Supplementary Figure 13 | Tunability at 50 GHz.** Electric-field-dependent real anisotropic permittivity $\epsilon_r$ of all measured films ($n$ = 6, 12, 16, 20, 24, 36).

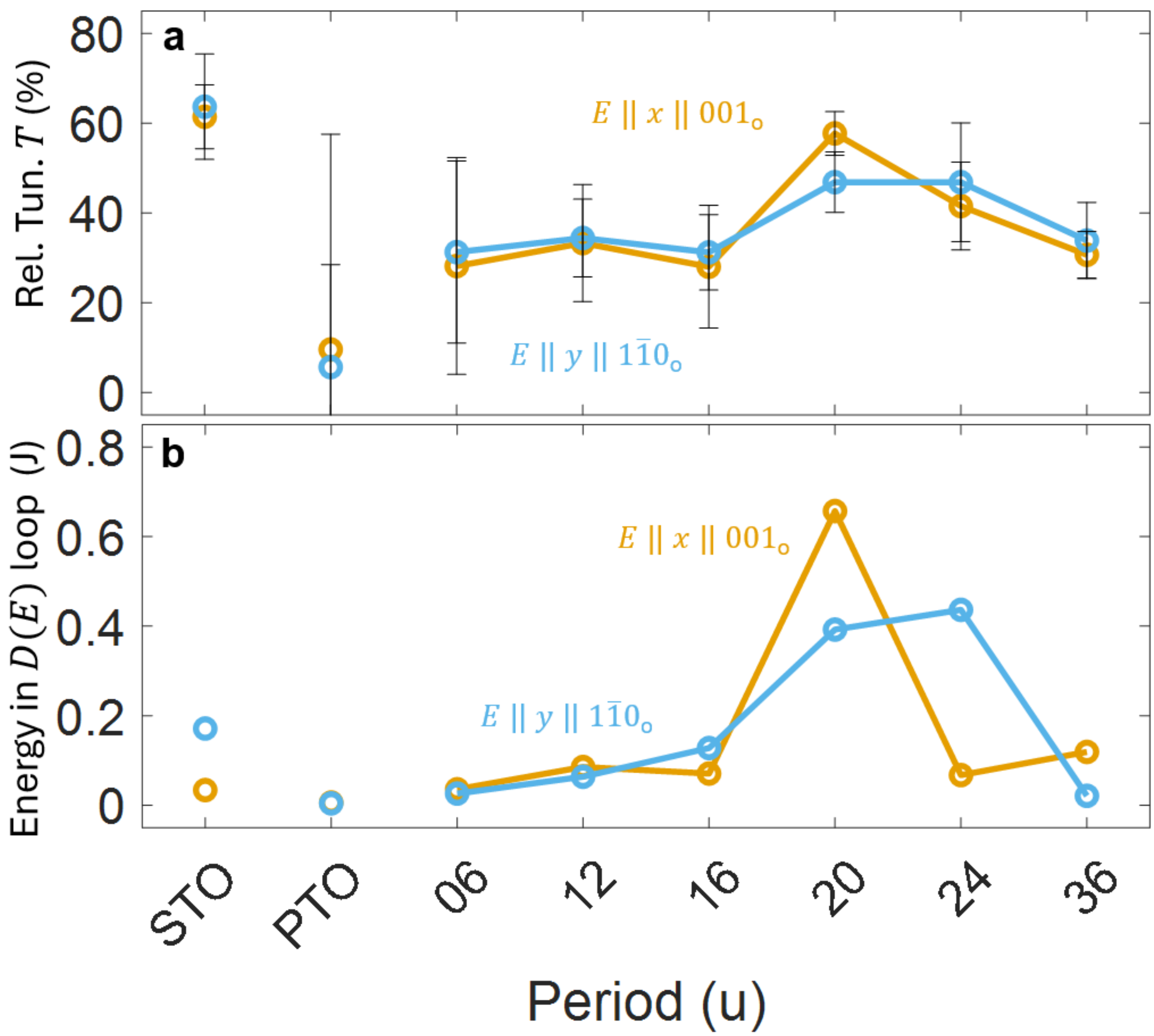

**Supplementary Figure 14 | Emergent tunability and ferroelectricity. a** Tunability for the different superlattice periodicities $n$ at 50 GHz under 100 kV/cm. We calculate the relative tunability $T$ via: $T = 1 - \epsilon'(100\,\mathrm{kV/cm})/\epsilon'_{\mathrm{max}}$, where $\epsilon'_{\mathrm{max}}$ is the maximum real permittivity over the complete voltage cycle and does not necessarily need to be at $E$ = 0 kV/cm. The vortex f film ($n$ = 20) shows a peak in comparison to the other superlattices but does not reach the tunability of the STO control film. **b** Energy in the $D(E)$ loop as measure for the ferroelectricity. The vortex film ($n$ = 20) shows emergent tunability in both measured orientations.

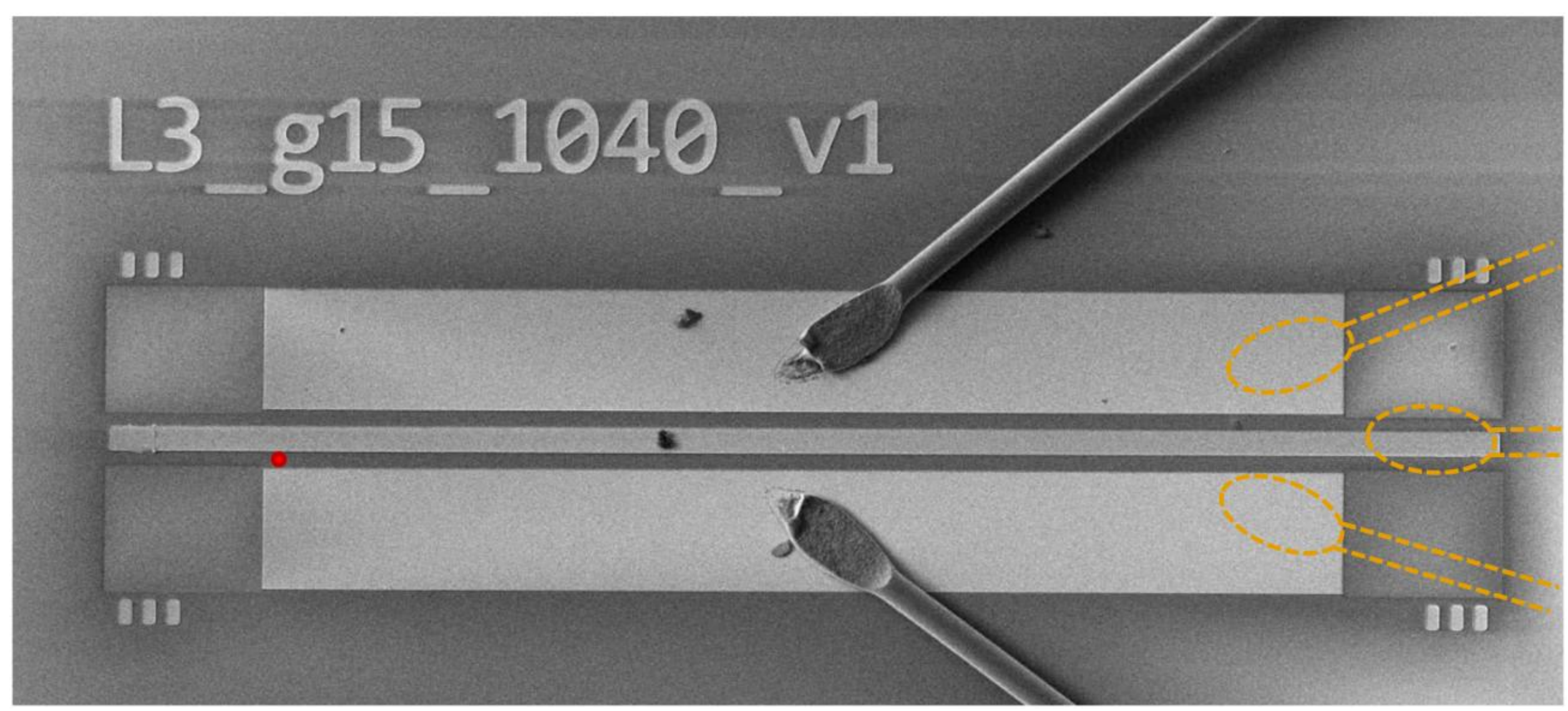

**Supplementary Figure 15 | Pulsed Laser Interferometry on-wafer setup.** Scanning Electron Microscopy (SEM) image of after Focused Ion Beam (FIB) etching of the landing pads to make room for wire bond feeds. The wire bonds visible in the image are only for grounding purposes during the FIB etching. The position of the wirebonds that are used during the PLI measurement are sketched into the image. The red dot depicts the approximate spot where the frequency sweep (**Fig. 4g**) was taken.

**Supplementary Table 1 | Phase-field parameters.** List of parameters used in the phase-field simulations for the $PbTiO_3$ and for $SrTiO_3$.

| Coefficient | $PbTiO_3$ [69] | $SrTiO_3$ [70] |
|---|---|---|
| $a_{11}$ | $3.8 \times 10^5 (T - T_c) \left(\frac{\mathrm{J}}{\mathrm{m}^3} \frac{\mathrm{m}^4}{\mathrm{C}^2}\right)$ <br> $T_c = 752\ \mathrm{K}$ | $7.50 \times 10^5\ T_s \left(\coth\left(\frac{T_s}{T}\right) - \coth\left(\frac{T_s}{T_c}\right)\right) \left(\frac{\mathrm{J}}{\mathrm{m}^3} \frac{\mathrm{m}^4}{\mathrm{C}^2}\right)$ <br> $T_c = 30\ \mathrm{K}$ <br> $T_s = 54\ \mathrm{K}$ |
| $a_{1111}$ | $-7.252 \times 10^7 \left(\frac{\mathrm{J}}{\mathrm{m}^3} \frac{\mathrm{m}^8}{\mathrm{C}^4}\right)$ | $1.70 \times 10^9 \left(\frac{\mathrm{J}}{\mathrm{m}^3} \frac{\mathrm{m}^8}{\mathrm{C}^4}\right)$ |
| $a_{1122}$ | $7.50 \times 10^8 \left(\frac{\mathrm{J}}{\mathrm{m}^3} \frac{\mathrm{m}^8}{\mathrm{C}^4}\right)$ | $3.92 \times 10^9 \left(\frac{\mathrm{J}}{\mathrm{m}^3} \frac{\mathrm{m}^8}{\mathrm{C}^4}\right)$ |
| $a_{111111}$ | $2.606 \times 10^8 \left(\frac{\mathrm{J}}{\mathrm{m}^3} \frac{\mathrm{m}^{12}}{\mathrm{C}^6}\right)$ | $0 \left(\frac{\mathrm{J}}{\mathrm{m}^3} \frac{\mathrm{m}^{12}}{\mathrm{C}^6}\right)$ |
| $a_{111122}$ | $6.10 \times 10^8 \left(\frac{\mathrm{J}}{\mathrm{m}^3} \frac{\mathrm{m}^{12}}{\mathrm{C}^6}\right)$ | $0 \left(\frac{\mathrm{J}}{\mathrm{m}^3} \frac{\mathrm{m}^{12}}{\mathrm{C}^6}\right)$ |
| $a_{112233}$ | $-3.70 \times 10^9 \left(\frac{\mathrm{J}}{\mathrm{m}^3} \frac{\mathrm{m}^{12}}{\mathrm{C}^6}\right)$ | $0 \left(\frac{\mathrm{J}}{\mathrm{m}^3} \frac{\mathrm{m}^{12}}{\mathrm{C}^6}\right)$ |
| $C_{11}$ | $180 \times 10^9$ (Pa) | $180 \times 10^9$ (Pa) |
| $C_{12}$ | $80 \times 10^9$(Pa) | $80 \times 10^9$(Pa) |
| $C_{44}$ | $110 \times 10^9$(Pa) | $110 \times 10^9$(Pa) |
| $Q_{11}$ | 0.089 ($\mathrm{m}^4/\mathrm{C}^2$) | 0.0457 ($\mathrm{m}^4/\mathrm{C}^2$) |
| $Q_{12}$ | −0.026 ($\mathrm{m}^4/\mathrm{C}^2$) | −0.0135 ($\mathrm{m}^4/\mathrm{C}^2$) |
| $Q_{44}$ | 0.034 ($\mathrm{m}^4/\mathrm{C}^2$) | 0.0096 ($\mathrm{m}^4/\mathrm{C}^2$) |
| $G_{11}$ | $1.04 \times 10^{-10}$ ($\mathrm{J\,m}^3/\mathrm{C}^2$) | $1.04 \times 10^{-10}$ ($\mathrm{J\,m}^3/\mathrm{C}^2$) |
| $G_{12}$ | 0 ($\mathrm{J\,m}^3/\mathrm{C}^2$) | 0 ($\mathrm{J\,m}^3/\mathrm{C}^2$) |
| $G_{44}$ | $0.52 \times 10^{-10}$ ($\mathrm{J\,m}^3/\mathrm{C}^2$) | $0.52 \times 10^{-10}$ ($\mathrm{J\,m}^3/\mathrm{C}^2$) |
| $\rho$ | $7.5 \times 10^3$($\mathrm{Kg/m}^3$) | $7.5 \times 10^3$($\mathrm{Kg/m}^3$) |
| $\beta$ | $6.0 \times 10^{-12}$(s) | $6.0 \times 10^{-12}$(s) |
| $\kappa^b$ | 20 (unitless) | 20 (unitless) |
| $\mu$ | $9.71 \times 10^{-17} \left(\frac{\mathrm{Kg}}{\mathrm{m}} \frac{\mathrm{m}^4}{\mathrm{C}^2}\right)$ | $9.71 \times 10^{-17} \left(\frac{\mathrm{Kg}}{\mathrm{m}} \frac{\mathrm{m}^4}{\mathrm{C}^2}\right)$ |
| $\gamma$ | $5.88 \times 10^{-6} \left(\frac{\mathrm{Kg}}{\mathrm{ms}} \frac{\mathrm{m}^4}{\mathrm{C}^2}\right)$ | $5.88 \times 10^{-6} \left(\frac{\mathrm{Kg}}{\mathrm{ms}} \frac{\mathrm{m}^4}{\mathrm{C}^2}\right)$ |

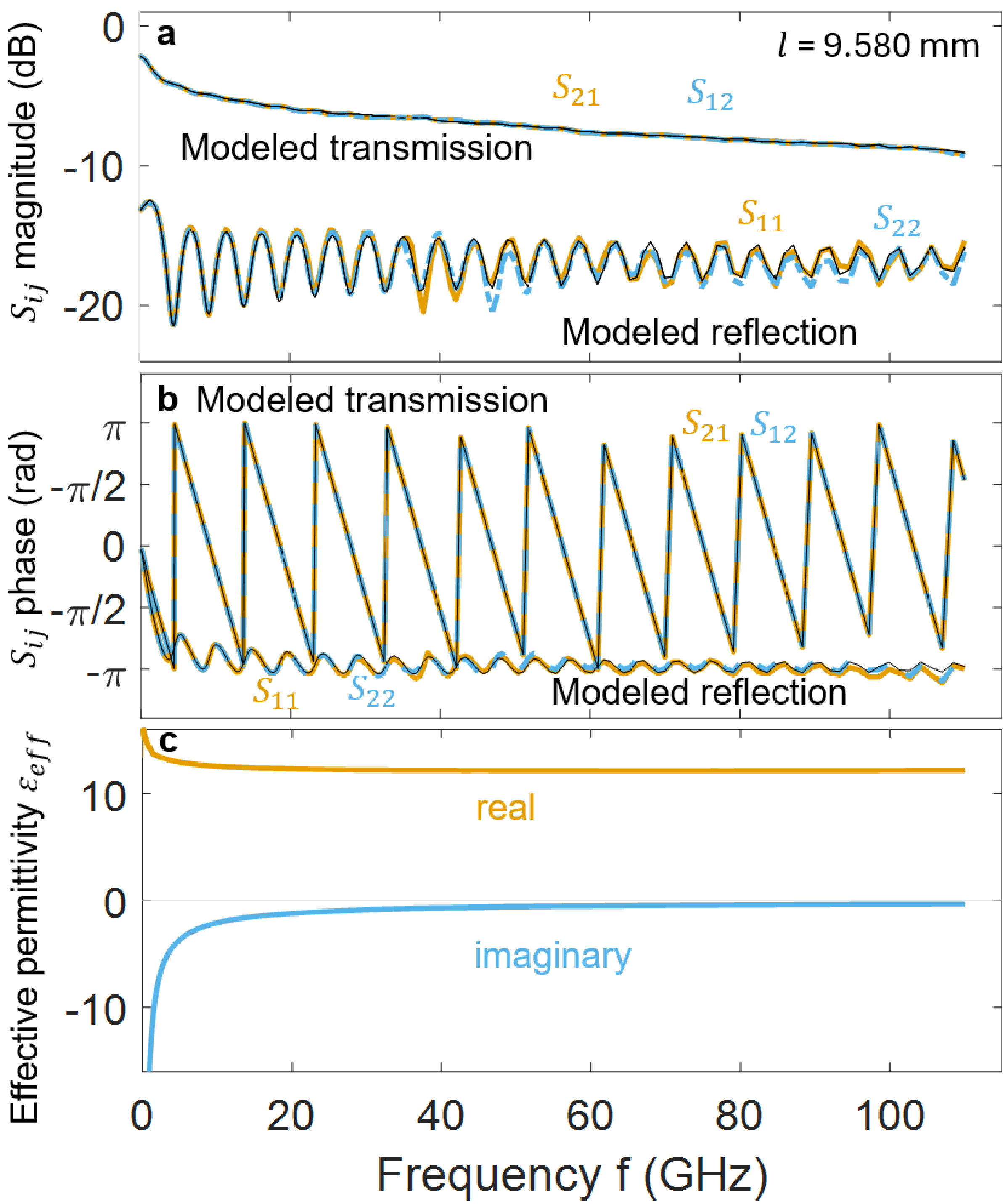

**Supplementary Figure 16 | Controlling the 1st-tier calibration on the LSAT substrate.** Comparison of calibrated and modeled S-parameters for the longest line in **a** magnitude and **b** phase, respectively. **c** Effective permittivity of the CPW transmission lines on the LSAT calibration wafer.

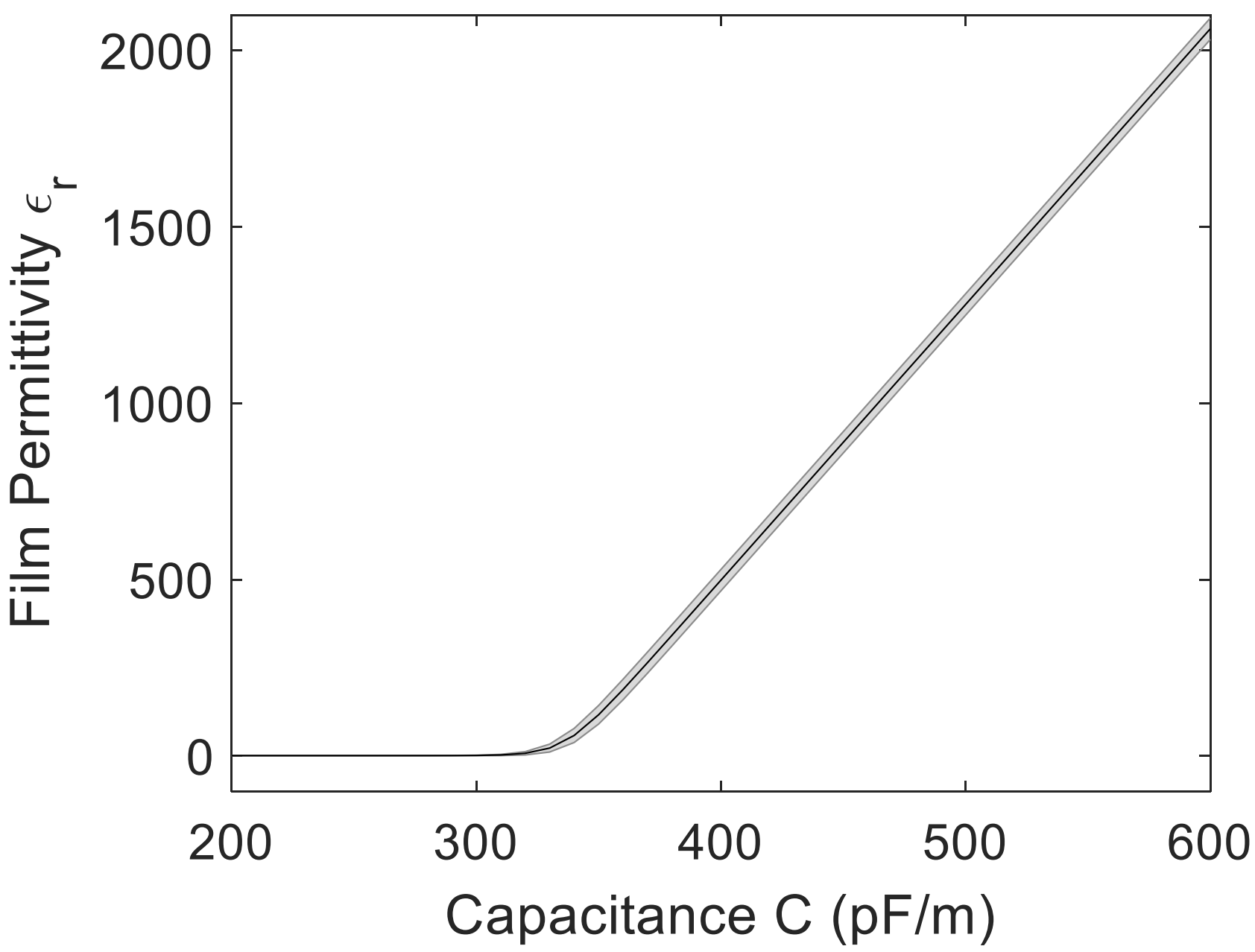

**Supplementary Figure 17 | Mapping Capacitance to Permittivity.** Example of a mapping function to obtain the film permittivity from capacitance per unit length of the transmission lines. The grey area is the standard uncertainty.

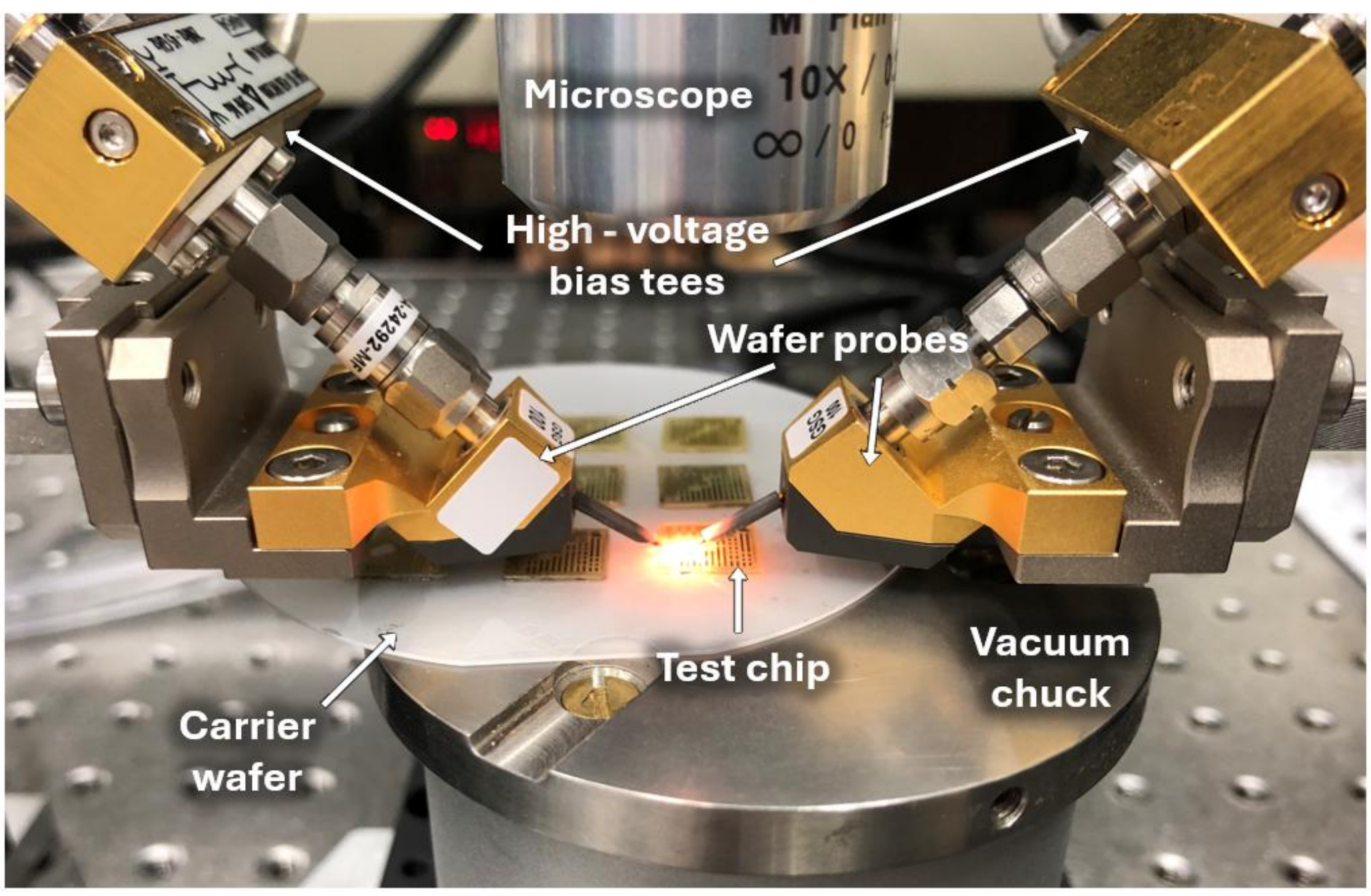

**Supplementary Figure 18 | Millimeter-wave setup.** Photo of the on-wafer millimeter-wave setup for tunability measurements up to 50 GHz and 150 V.

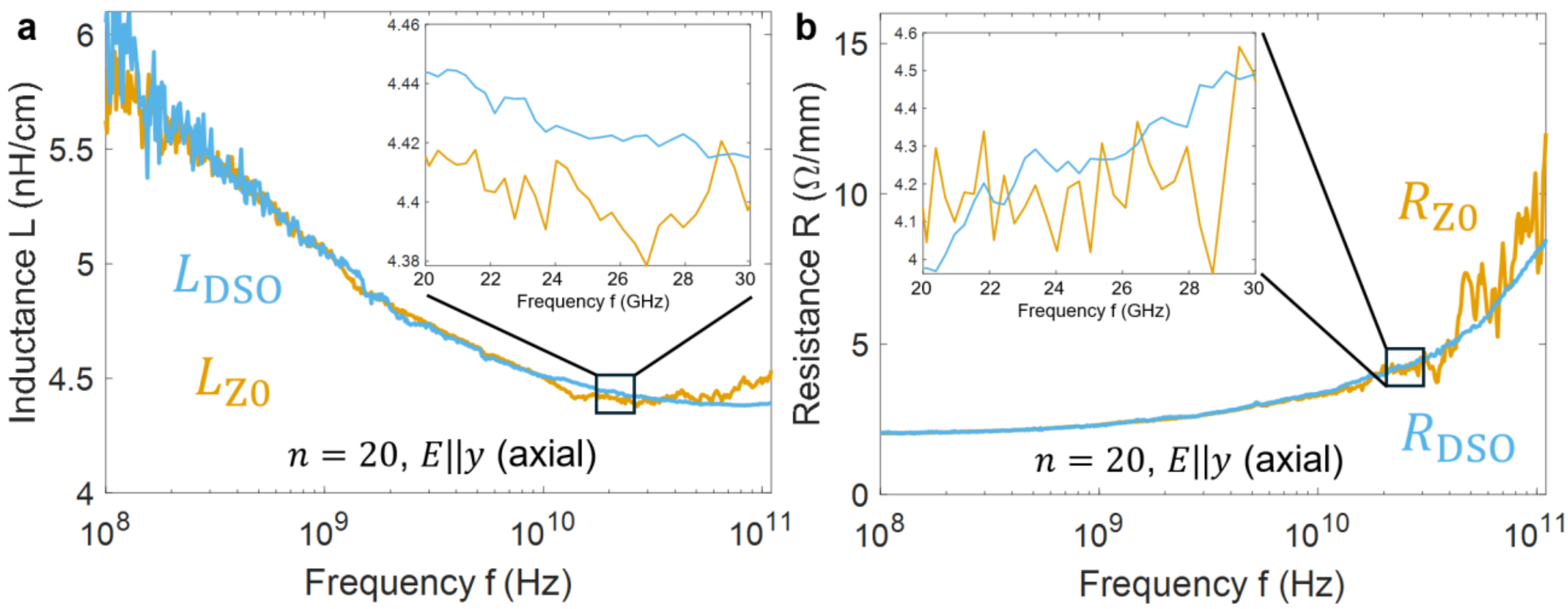

**Supplementary Figure 19 | Excluding magnetic origins of resonance.** Test excluding that the 25 GHz resonance in the $n$ = 20 film is of magnetic origin by allowing the circuit parameters $R$ and $L$ to vary freely ($Z_0$) instead of restricting them to the circuit parameters measured on the DSO substrate (DSO).

---

*Usage of commercial products herein is for information only; it does not imply recommendation or endorsement by NIST.